\documentstyle[a4wide,afd-gen,afd-eis,afd-orm,tr,url,biblio,times]{article}

\title{Conceptual Schema Optimisation --\\
       Database Optimisation before sliding down the Waterfall}
\author{H.A. Proper$^1$ and T.A. Halpin$^2$\\
        Department of Computer Science\\
        University of Queensland\\
        Brisbane\\
        Australia 4072\\
        E.Proper@acm.org}
\date{\Version}
\DoParSkip

   \input{epsf}
	 \def\Scale{0.5}
	 \def\epsfsize#.##2{\Scale#.}
   \def\Figurepath{.}

\begin{document}
   \maketitle
   \footnotetext[1]{Part of this work has
                 been supported by an Australian Research Council grant,
                 entitled: ``An expert system for improving complex 
                 database design''}
   \footnotetext[2]{Currently on leave at Asymetrix Corporation, Bellevue WA, USA}

   {\sc Published as:}
\begin{quote}
  H.A.~(Erik) {Proper} and T.A. {Halpin}. {Conceptual Schema Optimisation -- Database Optimisation before sliding down the Waterfall}. Technical report, Department of Computer Science, University of Queensland, Brisbane, Queensland, Australia, July 1995.
\end{quote}

   \begin{abstract}
   In this article we discuss an approach to database optimisation in which
   a conceptual schema is optimised by applying a sequence of 
   transformations.
   By performing these optimisations on the conceptual schema, a large
   part of the database optimisation can be done before actually 
   sliding down the software development waterfall.

   When optimising schemas, one would like to preserve some level of 
   equivalence between the schemas before and after a transformation.
   We distinguish between two classes of equivalence, one
   based on the mathematical semantics of the conceptual schemas, and
   one on conceptual preference by humans.

   As a medium for the schema transformations we use the universe of all
   (correct) conceptual schemas.
   A schema transformation process can then be seen as a journey (a schema-
   time worm) within this universe.
   The underlying theory is conveyed intuitively with sample transformations,
   and formalised within the framework of Object-Role Modelling. 
   A metalanguage is introduced for the specification of transformations, and
   more importantly their semantics. While the discussion focusses on the 
   data perspective, the approach has a high level of generality and is 
   extensible to process and behaviour perspectives. 
\end{abstract}

   \section{Introduction}
\SLabel{section}{intro}

Modern approaches to information system development usually start out by 
modelling a universe of discourse in terms of a conceptual schema, using the
notation of a conceptual modelling method such as Enhanced Entity 
Relationship (EER) modelling or Object-Role Modelling (ORM). 
The design of this schema is ideally guided by some kind of conceptual
schema design procedure. For many years, ORM has featured well developed design
procedures (\cite{Book:90:Wintraecken:NIAM},
\cite{Book:94:Halpin:ORM}), and some variants of EER now include design techniques (e.g.\ \cite{Book:94:Elmasri:DBFundamentals}, 
\cite{Book:92:Batini:ConceptualDBDesign}). 
Most modern Object-Oriented (OO) approaches also include a design
procedure (\cite{Book:91:Rumbaugh:OO}, \cite{Book:90:Coad:OO},
\cite{Book:94:Kristen:KISS}).

After a conceptual schema of a given universe of discourse is finalised, it 
can be mapped to a database schema for the actual implementation.
This database schema might include a set of tables for a relational 
database management system, or a set of classes for an object-oriented 
database management system.
During the mapping from a conceptual schema to a database schema,
optimisation issues usually come into play,
since one would like the resulting database schema to have
a good performance in the given context of the application.
In practice this usually means that either during the mapping 
or afterwards, optimisations of the database schema
are made.
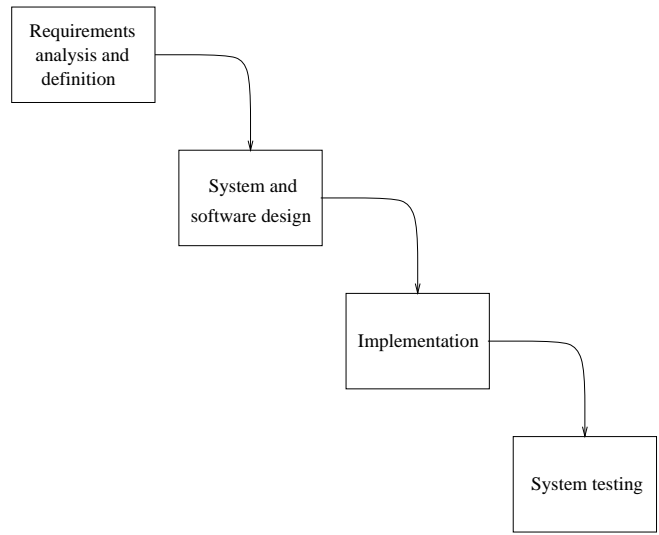
\EpsfFig[\Waterfall]{The simple waterfall view of the software development life-cycle}

In this article we are concerned with an approach that allows us to do schema
optimisations on the conceptual schema, i.e.\ before sliding down
the waterfall (see \SRef{\Waterfall}).
We realise that the waterfall model only provides a simplistic view
on an information system development life cycle, and that in practice 
the waterfall contains a number of sub-cycles.
However, it does provide a good framework to discuss our ideas, 
although the ideas are independent of the actual waterfall model.

Although not all database schema optimisations can be done at 
this higher level, 
many optimisations can indeed be done
at this level, offering a number of advantages.
For example, conceptual schemas provide a clearer, human-oriented picture 
of the universe of discourse, so user participation during the optimisation 
process is more viable, where different schema alternatives are discussed.
Moreover, at the conceptual level, schema transformations may be studied in a 
way independent of the implementation platform (relational, object-oriented, 
hierarchical, \ldots). 

To provide a better context for the contributions made in
this article, we now describe the information
system design process itself as a schema transformation process.
Viewing the whole process of transforming a draft conceptual schema to
the final database schema as a sequence of schema transformations, is
an idea that is closely related to the idea of program derivation by a 
sequence of transformations 
(\cite{Article:79:Bauer:Trans}, \cite{Article:83:Partsch:Trans},
\cite{Article:89:Partsch:ProgConstr}).
In our view, the following kinds of schema transformations 
(\cite{Article:90:Halpin:Wise}) may be distinguished during a database
modelling process:
\begin{enumerate}
   \item Conceptual schema draft
   \item Conceptual schema refinement
   \item Conceptual schema optimisation
   \item Conceptual to database schema mapping
   \item Database schema optimisation
\end{enumerate}
We now briefly discuss each of these phases.
When a conceptual schema is oginally drafted, it typically goes through a
series by transformations to improve the current design, in accordance with 
the design procedure. Most of these transformations do not maintain 
equivalence. Extra information may be added, and unwanted details may be removed.

Before or after a universe of discourse is completely captured in a conceptual
schema, it might be discovered that certain parts of this conceptual schema 
have alternative representations.
One of these alternatives will usually be considered preferable in terms of the
way that the user(s) wish to think about the application. 
So after the initial conceptual schema has been developed, a series of 
transformations might be performed that lead to a schema that is 
a preferred view of the universe of discourse.
This class of transformations normally has to preserve the (mathematical) 
equivalence of the schemas as they deal with alternatives describing one 
and the same universe of discourse.
As an example, consider the ORM schemas in \SRef{\ExTrans}. 
Here object types are shown as named ellipses with their identification schemes
in parentheses; roles are shown as boxes attached to the object types that play them; 
predicates are depicted as named sequences of roles; constraints on values are listed
in braces; and arrow-tipped bars denote uniqueness constraints on roles. 
Here the uniqueness constraints are the weakest possible (e.g.\ a lecturer may teach
many students, and a student may be taught by many lecturers).    
The two ORM/NIAM schemas depicted are mathematically equivalent
(for a proof of this, refer to \cite{PhdThesis:89:Halpin:NIAMForm}).
These two alternatives however might not be equally preferable ways 
for the user(s) to think about the application. 
The schemas in the example are modelled using the Object-Role Modelling
(ORM) technique (\cite{Article:88:Leung:NIAM}, \cite{Report:91:Hofstede:PSM}, 
\cite{Book:90:Wintraecken:NIAM}, \cite{Book:94:Halpin:ORM}).
Similar examples for EER or OO models could be given.

Once the user has selected the preferred conceptual schema, this should be 
used in any later conceptual queries on the implemented information system, 
assuming the availablity of an appropriate  
conceptual query language (see e.g.\ 
\cite{Report:82:Meersman:RIDL}, \cite{Report:91:Hofstede:LISA-D},
\cite{Report:93:Proper:DisclSch}) 
Before mapping this preferred conceptual schema to the target database schema
we may wish to perform some further optimisations to this conceptual schema,
under the covers. 
As an example of such a transformation, again, consider \SRef{\ExTrans}.
Depending on the {\em data profile} and {\em access profile}, either 
schema fragment may be more efficient to implement than the other.
More examples of such conceptual schema optimisations can for instance be found
in \cite{Article:90:Halpin:SchemaOpt}, \cite{Article:91:Halpin:SchTransf},
\cite{Article:92:Halpin:SchemaOptim}, \cite{Book:92:Batini:ConceptualDBDesign}.

To prevent confusion in the remainder of this article, we will  
use the term {\em data schema} when we refer to an ORM schema in general, 
be it a preferred or only a draft conceptual representation of a universe of discourse.
The term {\em conceptual schema}
will typically be used for the preferred conceptual view.
The term {\em database schema} is reserved for the target schema in the chosen
database management system, which could be a relational, object-oriented,
or hierarchical system.
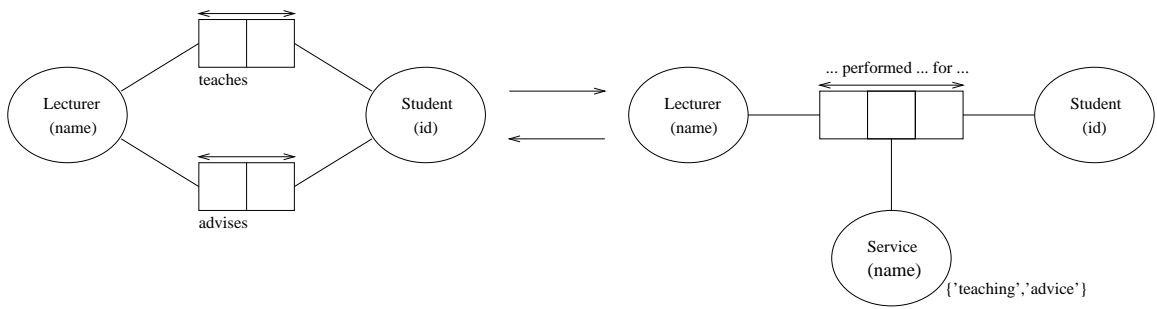
\EpsfFiG[\ExTrans]{Example equivalence preserving schema transformation}

When a data schema is mapped to a database schema, in general, a 
mapping algorithm is used that tries to find a (nearly) optimal 
representation.
A wide range of algorithms for this purpose exists, for instance:
\cite{Article:86:Berman:DataTrans},
\cite{Article:87:Shoval:ADDS},
\cite{Report:91:Bommel:CSTransform}, \cite{Article:93:Rishe:DataTrans}, 
\cite{Article:93:McCormack:Mapping}, \cite{Book:94:Halpin:ORM},  
\cite{Article:94:Bommel:ImplSel}, and
\cite{PhdThesis:94:Ritson:ORMMap}.
For the reverse process, reverse engineering, also a wide range of
strategies and algorithms exists 
(e.g. \cite{Article:91:Kalman:RevEng},
\cite{Article:92:Fonkam:RevEng},
\cite{Article:93:Shoval:Reverse},
\cite{Article:94:Chiang:RevEng}).
Note that reverse engineering may also be regarded
as a sequence of schema transformations (basically the reverse of those used 
for conceptual schema to internal schema mapping).

Once a data schema has been represented as a database schema, this 
schema may sometimes be optimised further using transformations 
of the database schemas and adding indexes.
Some of these transformations are discussed in e.g.
\cite{PhdThesis:93:DeTroyer:DataTrans}, 
\cite{Article:86:Kobayashi:BinRelTransform}, and 
\cite{Book:92:Batini:ConceptualDBDesign}.

The discussed five classes of transformations operate either on a 
data schema, or an internal schema of a given implementation platform.
The interplay of these transformations is illustrated in \SRef{\CSDP}.
The modelling process up until the start of the database schema mapping
can be regarded as a journey through the universe of data (ORM) schemas.
This article focusses on schema transformations involved in
classes 2 and 3.
Furthermore, we limit ourselves to ORM models.
This latter limitation, however, is not a strong one as the ORM modelling
technique is general enough to cater for ER based data schemas as well 
(\cite{Report:94:Brouwer:ORMKernel},
\cite{Report:94:Halpin:ORMPoly}).
In \cite{Article:94:Campbell:Abstraction},
\cite{Article:94:Campbell:AbstractionORM} and
\cite{Report:95:Proper:CDMKernel} it is shown how ER can be 
regarded as an abstraction from ORM schemas.

Research is also underway (together with the authors of 
\cite{Report:91:Bommel:CSTransform}) to find an apt schema language 
to describe both data schemas as well as database schemas.
This would allow us to describe the entire modelling process of an information
system's data(base) schema within one modelling language.
A likely candidate for this aim is the tree representation of ORM
schemas as introduced in \cite{Report:91:Bommel:CSTransform}.
This technique is not a replacement of ORM, but rather an extension
which allows the representation of candidate internal representations 
besides the normal data schema.
These candidate internal representations have a one to one correspondence 
to relational models, NF$^2$ models, 
network, hierarchical and O$^2$ models.
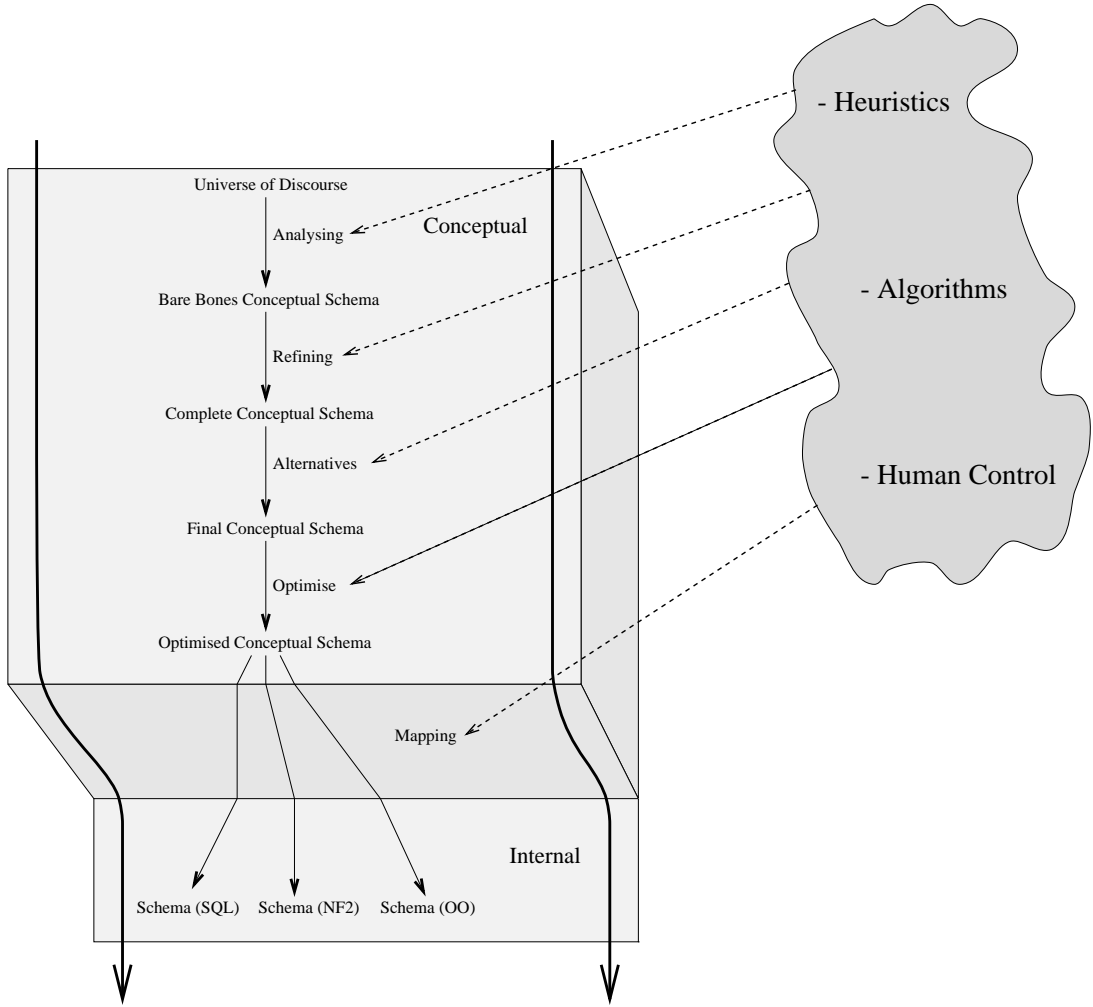
{\def\Scale{0.6} \EpsfFig[\CSDP]{Single schema language}}

The structure of the remainder of this article is as follows.
Although we also consider schema transformations which do not
maintain equivalence, it is essential to define exactly what we regard
as schema equivalence.
Therefore, in \SRef{SchemaEquiv} this notion is discussed in more detail.
A more elaborate discussion of example transformations and their applications
is given in \SRef{ExTransforms}.
As stated before, the design process of a data schema can be seen 
as making a journey through the universe of data schemas.
In \SRef{ORMU} we therefore define the universe of ORM data schemas,
while the notion of a schema version representing a state of the data
schema during the design process is discussed in \SRef{ORMV}. 
In this article we focus on the syntax of ORM models in the context
of this universe.
The semantics has been discussed in detail elsewhere.
A language to define the schema transformation is introduced in
\SRef{TransLang}.
Before concluding, we define in \SRef{ApplyTrans} three ways in 
which to apply the schema transformations to an existing data schema.

   \section{Equivalence of Schemas}
\SLabel{section}{SchemaEquiv}

In the context of schema transformations two notions of schema equivalence
are important.
The first notion of equivalence is mathematical equivalence.
Two schemas are mathematically equivalent if they define isomorphic
state spaces.
The second notion of equivalence tries to capture the conceptual
quality of a conceptual schema.
When there are two mathematically equivalent alternative schemas
for the same universe of discourse, one alternative may still be a better
representation of the domain than the other.
For example, consider the two schemas shown in \SRef{\ExTrans}.
For a given universe of discourse, one of these two may be more 
`natural'.
We first discuss the notion of mathematical equivalence.

\subsection{Mathematical equivalence}
We start by discussing three existing approaches to defining schema 
equivalence and their inter-relationships.
In the remainder of this article, whenever we refer to {\em equivalence}
without using the prefix {\em mathematical} or {\em conceptual}, we implicitly 
refer to {\em mathematical equivalence}.

\subsubsection{Set based equivalence}
A data schema can be seen as an intensional specification of a set of valid 
populations. 
This is what we call the semantics of the schema.
Even in databases which maintain the history 
(\cite{Article:92:Roddick:TimeSurvey}) of the population this holds.
In such cases, the schema semantics of the data schema is the set of valid
populations at each point in time. 
If the schema does not evolve in the course of time, then the history of the
database takes place within this fixed set of valid populations.
How this view needs to be adapted in the case of evolving schemas is
discussed in \cite{Report:92:Proper:EvolvAM} and 
\cite{Report:93:Proper:EvolvPSM}.

Usually a population of a data schema $\Schema$ is modelled as a function:
\[ p: \Types \Func \Powerset(\UniDom) \]
where $\Types$ is the set of (populatable) types defined in $\Schema$ and 
$\UniDom$ is some domain of instances. 
The state space of a data schema $\Schema$ can then be defined as: 
\[ 
   \Cal{S}(\Schema) ~\Eq~ 
   \Set{p:\Types \Func \UniDom}{\IsPop(\Schema,p)} 
\]
where $\IsPop$ is a predicate that determines whether $p$ is a proper 
population of $\Schema$. 
Two data schemas are now equivalent iff there exists a bijection between 
their state spaces (\cite{Report:92:Hofstede:SchemaEquiv}), which is 
equivalent to saying that the state spaces are equally sized (could be 
infinite):
\[ 
   \Schema_1 \equiv \Schema_2 ~\Eq~
   \Ex{h}{
     h \mbox{~is a bijection~} h:\Cal{S}(\Schema_1) \Func \Cal{S}(\Schema_2)
   } 
\]
A direct result of the above two definitions is:
\begin{corollary}
   If $h$ is a bijection $h:\Cal{S}(\Schema_1) \Func \Cal{S}(\Schema_2)$,
   then: 
   \[ \IsPop(\Schema_1,p) \iff \IsPop(\Schema_2,h(p)) \]
\end{corollary}
Using this definition, it indeed becomes provable that every ORM schema with a 
finite ($n$) set of populations is equivalent to a schema of the form 
depicted in \SRef{\GenOrm} (see also  \cite{Report:92:Hofstede:SchemaEquiv}).
Note that the \SF{\#=1} in this figure indicates that the population of
value type \SF{NatNo} only contains one instance, and \SF{\{1..n\}}
limits the instances of \SF{NatNo} to the interval $1$ to $n$.
The proof that each ORM schema is equivalent to this schema is based on 
the fact that the size of the populations is 
finite, and therefore there exists a bijection between the state space of the 
ORM schema and a subset of the natural numbers.
The schema depicted in \SRef{\GenOrm} simply corresponds to a data 
schema where each population corresponds to {\em one} natural number.
We realise that this is an unlikely outcome of any schema optimisation 
process as it moves the optimisation difficulty from storing information to 
decoding the information.
However one may argue that this schema does have a correspondence to reality, 
as all populations are stored on a hard disk as a sequence of bits (grouped
in bytes) which can be seen as one large natural number.
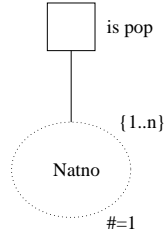
\EpsfFig[\GenOrm]{Most general schema}

\subsubsection{Logic based equivalence}
In \cite{PhdThesis:93:DeTroyer:DataTrans} and 
\cite{Article:86:Kobayashi:SchemaEquivalence} a logic 
based notion of schema equivalence is introduced. 
In this approach a schema is interpreted as a logic 
signature with an associated set of basic axioms (the constraints). 
The notion of equivalence is then defined as:
\[ \Schema_1 \equiv \Schema_2 ~\Eq~
   \Ex{h}{
     h \mbox{~is a bijection~} h:\Msy{I}(\Schema_1) \Func \Msy{I}(\Schema_2)
   }
\]
where $\Msy{I}(\Schema)$ is the set of valid interpretations of data schema
$\Schema$.
This exactly corresponds to the above notion of equivalence, since 
$\Msy{I}(\Schema)$ corresponds directly to the notion of state space.

\subsubsection{Contextual equivalence}
The notion of equivalence as defined in \cite{PhdThesis:89:Halpin:NIAMForm} 
is based on a more pragmatic approach. 
The above discussed notions of equivalence are not concerned with finding a 
proof of the equivalence, whereas the approach described in 
\cite{PhdThesis:89:Halpin:NIAMForm} is. 
Every schema $\Schema$ can be seen as a set of logic formulae 
$\Msy{L}(\Schema)$ describing the structure and the constraints of the 
conceptual schema\footnote{One might argue that
not all schemas of all data modelling techniques can be expressed in terms
of a set of First Order Predicate Calculus formulae, but for most techniques
that are in actual use this can indeed be done}.
Given two of such sets ($\Msy{L}(\Schema_1)$, $\Msy{L}(\Schema_2)$) the 
question of equivalence of schemes $\Schema_1$ and $\Schema_2$ then 
corresponds to the question whether $\Msy{L}(\Schema_1)$ is provable from 
$\Msy{L}(\Schema_2)$ and vice versa:
\[ \Msy{L}(\Schema_1) \iff \Msy{L}(\Schema_2) \]
However, in $\Schema_1$ one may have introduced other predicates 
(other names, other arities) than are present in $\Schema_2$ (and vice versa).
Therefore, the need arises to provide some extra formulae to define a 
translation between these predicates. 
Therefore, the equivalence of $\Schema_1$ and $\Schema_2$ is
defined as:
\[ \Schema_1 \equiv \Schema_2 ~\Eq~  
   \Msy{L}(\Schema_1) \land D_1  \iff  \Msy{L}(\Schema_2) \land D_2 
\]
where $D_1$ and $D_2$ are the formulae providing the translation between the 
two sets of predicates. 
This leads to the notion of {\em contextual equivalence}; the set of formulas 
$D_1$ and $D_2$ provide the context of the equivalence. 
Both $D_1$ and $D_2$ are conjunctions of formulae such that the predicates 
(and functions) provided in $\Schema_2$ are defined in terms of the ones 
provided in $\Schema_1$ and vice versa. 
The theories $\Msy{L}(\Schema_1) \land D_1$ and $\Msy{L}(\Schema_2) \land D_2$ 
should form {\em conservative extensions} (see e.g. 
\cite{Book:77:Chang:ModelTheory}) of $\Msy{L}(\Schema_1)$ and 
$\Msy{L}(\Schema_2)$. 
Proving the logical equivalence of $\Msy{L}(\Schema_1) \land D_1$ and 
$\Msy{L}(\Schema_2) \land D_2$ is equivalent to proving:
\[ \Msy{I}(\Msy{L}(\Schema_1) \land D_1)  = 
   \Msy{I}(\Msy{L}(\Schema_2) \land D_2) \]
Using {\em contextual equivalence}, a number of schema equivalence problems 
become provable using a theorem prover 
(\cite{PhdThesis:93:Bloesch:TheoremProver}). 
Note that in general not all schema equivalence problems are automatically 
provable, or even decidable.

From the translations between predicates provided by $D_1$ and $D_2$ a
bijection $h$ as used in the above discussed approaches can be derived.
All schemas that can be proven to be equivalent using the approach
based on {\em contextual equivalence} are therefore equivalent in the
sense of the first two approaches.
However, the reverse does not necessarily hold. 
For example, consider the schemas depicted in \SRef{\Rename}.
The two schemas are isomorphic, but we cannot prove contextual
equivalence of the two schemas since this would lead to a naming problem
in the conservative extensions.
Nevertheless, in these cases equivalence can be proven by extending the name 
space of roles and types.
A name $n$ used in the left schema could be extended to $1.n$, while
a name $n$ used in the right schema could be extended to $2.n$.
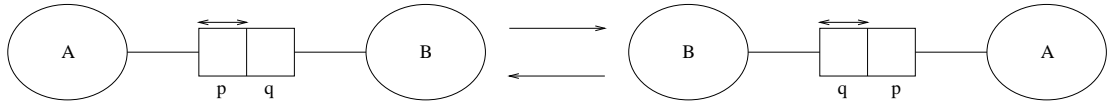
\EpsfFig[\Rename]{Renaming roles and types}

\subsubsection{Substitution property}
In traditional mathematics, if $X$ and $Y$ are expressions yielding a natural 
number and we have proven that $X = Y$, then we also know that 
$X + 1 = Y + 1$ since we are allowed to replace $X$ by $Y$.
This is an example of the substitution property.

Analogously we would like to have a substitution property for schemas.
Once we have proven the equivalence of two schemas $\Schema_1$ and
$\Schema_2$, we would like to be able to conclude the equivalence
of two schemas $\Schema'_1$ and $\Schema'_2$ if all in which they
differ from $\Schema_1$ and $\Schema_2$ respectively is a set of schema
components $X$.
Without such a property, it is impossible to build a general pool of equivalence
preserving transformations, as each time such a transformation is
applied the equivalence of the schemas would have to be re-proven.
Formally, we would like to have the following property:
\begin{theorem}(substitution theorem)
   Let $\Schema_1$ and $\Schema_2$ be schemas such that 
   $\Schema_1 \equiv \Schema_2$ and $X$ is a set of schema components. 

   If $\Schema'_1$ follows from $\Schema_1$ by adding components $X$, and
   similarly $\Schema'_2$ from $\Schema_2$ by adding components $X$, and
   $\Schema'_1$, $\Schema'_2$ are proper schemas, then:
   \[ \Schema'_1 \equiv \Schema'_2 \]
\end{theorem} 
However, the substitution property does not generally hold for
data schemas.
For example, in the case of \SRef{\Rename}, suppose we add the schema component
$\SF{Frequency}(q,1..3)$; which is a textual representation of a frequency
constraint on the role $q$. 
Adding this component to the two equivalent schemas results in two 
non-equivalent schemas!
The problem in this case is the fact that the frequency constraint
refers to the role $q$, which has a different semantics in both schemas.

The solution we propose is to require the schemas to provide the contextual
equivalence themselves, i.e.\ as derived types with proper derivation
rules.
Most modern data modelling techniques allow for the specification of
derivation rules.
We therefore introduce the notion of {\em direct equivalence} as:
\[ 
   \Schema_1 \equiv \Schema_2 ~\Eq~ 
   \Msy{L}(\Schema_1) \iff \Msy{L}(\Schema_2) 
\]
which equates to:
\[
   \Schema_1 \equiv \Schema_2 ~\Eq~ 
   \Msy{I}(\Msy{L}(\Schema_1)) = \Msy{I}(\Msy{L}(\Schema_2)) 
\]
Figure \ref{\Rename} is clearly not directly equivalent. As an example of a
direct equivalence, consider \SRef{\DirEqEx}, which represents the same 
universe of discourse as \SRef{\ExTrans}.

It is not hard to see that the substitution property holds in the case
of direct equivalence.
Any components added to $\Schema_1$ and $\Schema_2$ now have exactly
the same semantics, so if $\Schema'_1$ and $\Schema'_2$ follow from
the original schemas by adding components $X$ we have:
\[ \Msy{L}(\Schema_1) \SetMinus \Msy{L}(\Schema'_1) =
   \Msy{L}(\Schema_2) \SetMinus \Msy{L}(\Schema'_2) \]
Since $\Msy{L}(\Schema_1) \iff \Msy{L}(\Schema_2)$ we therefore have:
$\Msy{L}(\Schema'_1) \iff \Msy{L}(\Schema'_2)$.
\EpsfFig[\DirEqEx]{Example of direct equivalent schemas}

\subsubsection{Strengthening a schema}
A schema $\Schema_2$ is stronger than a $\Schema_1$ if each 
interpretation of $\Schema_2$ is an interpretation of $\Schema_1$,
so when:
\[ \Msy{I}(\Msy{L}(\Schema_1)) \supseteq \Msy{I}(\Msy{L}(\Schema_2)) \]
This leads to the following definition of a stronger schema:
\[ \Schema_1 \Stronger \Schema_2 ~\Eq~ 
   \Msy{L}(\Schema_1) \Leftarrow \Msy{L}(\Schema_2) \]
So when $\Schema_1 \Stronger \Schema_2$ we say that $\Schema_2$ is
stronger than $\Schema_1$.
In the next section we see an example of a schema strengthening
transformation.

\subsection{Conceptual equivalence}
\SLabel{subsection}{ConcEq}

The second class of schema equivalence is based on the relationship between
conceptual schemas and a universe of discourse.
Most ORM/NIAM based modelling techniques are based on the presumption that
when modelling a universe of discourse one actually constructs a grammar
of the communication taking place within this universe of discourse.
These modelling techniques therefore usually start out
from a set of sample sentences describing the universe of discourse.
These latter sentences are generally elicited from {\em domain
experts}, in accordance with the following postulate:
\begin{quote} \em
   Domain experts can fully describe their universe of discourse using (semi)
   natural language; which can sometimes be done in the form of a complete set
   of significant examples.
\end{quote}
This postulate may seem simple, but it is a crucial base of an increasing
number of modern approaches to conceptual modelling
(\cite{Article:89:Nijssen:BaseAxiom}, 
 \cite{Book:90:Coad:OO},
 \cite{Book:90:Wintraecken:NIAM},
 \cite{Book:94:Halpin:ORM}, 
 \cite{Book:94:Kristen:KISS}).
In principle, some schema features can never have a set of significant examples
(e.g. subtype definitions), but small significant example sets are easy to
generate for most constraints. The set of examples is usually gathered 
by using the so-called telephone paradigm 
(\cite{Book:94:Halpin:ORM}):
\begin{quote} \em
   Explain your observations to a non-expert via a telephone.
\end{quote}
The set of sentences that follows from this exercise defines a 
language; the {\em domain expert language} (\cite{Report:94:Hofstede:Grammar}).
The underlying grammar is referred to as $\Cal{G}\Sub{\ExpertLang}$.
The aim of a conceptual modelling process can now be described as
(\cite{Report:94:Hofstede:Grammar}):
\begin{quote} \em
   Find, within a certain class of sufficiently efficient computable grammars,
   a grammar which best approximates $\Cal{G}\Sub{\ExpertLang}$.
\end{quote}
The conceptual schema which follows from a conceptual design procedure in
itself defines a grammar $\Cal{G}\Sub{\Schema}$.
All correct ORM/NIAM/ER schemas essentially define such 
a computable grammar $\Cal{G}\Sub{\ExpertLang}$.
Therefore, the aim of the conceptual modelling process can be
re-formulated as:
\begin{quote} \em
   Find, a conceptual schema $\Schema$ such that the associated 
   grammar $\Cal{G}\Sub{\Schema}$ best approximates 
   $\Cal{G}\Sub{\ExpertLang}$.
\end{quote}
Please note that for $\Schema$ to be a correct schema certain well-formedness
criteria must be met, including restrictions on the verbalisations of the
fact types from the universe of discourse (\cite{Book:94:Halpin:ORM}).
For example, ORM requires the verbalised fact types to be elementary, i.e.\
the resulting facts should not be splittable into fact types of lower arity.

If we would have a formal notion of a distance between grammars:
\[ \SF{Distance}(\Cal{G}\Sub{\Schema},\Cal{G}\Sub{\ExpertLang}) = \delta \]
then we would be able to decide between two equivalent schema alternatives
for a given universe of discourse.
Let $\Schema_1$ and $\Schema_2$ be two equivalent schema alternatives
for a universe of discourse with expert language $\Cal{G}\Sub{\ExpertLang}$,
then $\Schema_1$ is a more {\em natural} description of the
universe of discourse if:
\[ \SF{Distance}(\Cal{G}\Sub{\Schema_1},\Cal{G}\Sub{\ExpertLang}) < 
   \SF{Distance}(\Cal{G}\Sub{\Schema_2},\Cal{G}\Sub{\ExpertLang}) \]
The schemas are conceptually equivalent when they have the same distance
to the expert language:
\[ \SF{Distance}(\Cal{G}\Sub{\Schema_1},\Cal{G}\Sub{\ExpertLang}) =
   \SF{Distance}(\Cal{G}\Sub{\Schema_2},\Cal{G}\Sub{\ExpertLang}) \]
Verifying such equivalences formally will most likely remain hard, if
not impossible.
The key problem being the fact that determining the equivalence between
grammars is known to be a hard problem.
Nevertheless, this definition of {\em naturalness} of a conceptual
schema with respect to a given universe of discourse does allow us to make
better design decisions in a conceptual schema design procedure when
considering alternative schemas.
 
   \section{Example Transformations}
\SLabel{section}{ExTransforms}

In this section we discuss some example applications of schema transformations
followed by their underlying schema transformations.
For a more complete treatise of such transformations, the reader is referred to
\cite{PhdThesis:89:Halpin:NIAMForm}, and \cite{Book:94:Halpin:ORM}.
In \cite{PhdThesis:89:Halpin:NIAMForm} proofs of equivalences
can be found as well.

\subsection{A simple schema transformation}
As a first simple example, consider the medical report shown below. 
Here a tick in the appropriate column indicates that a patient smokes or
drinks. 
Since both these ``vices'' can impair health, doctors are often
interested in this information. 
\begin{center}
   \begin{tabular}{|l|l|l|l|}
      \hline
      Patient\# & Patient name & Smoker? & Drinker?\\
      \hline \hline
      1001 & Adams, A   &       & \tick \\
      1002 & Bloggs, F  & \tick & \tick \\
      1003 & Collins, T &       & \\
      \hline
   \end{tabular}
\end{center}
Figure \ref{\HospitalA} shows one conceptual schema for this universe 
of discourse, together with the sample population. 
The black dot is a mandatory role constraint (each patient has a name). 
Here two optional unaries are used for the smoker-drinker facts. 
Instead of using unaries, we may model the smoker-drinker facts using
two functional binaries: 
\SF{Patient has SmokerStatus}; \SF{Patient has DrinkerStatus}. 
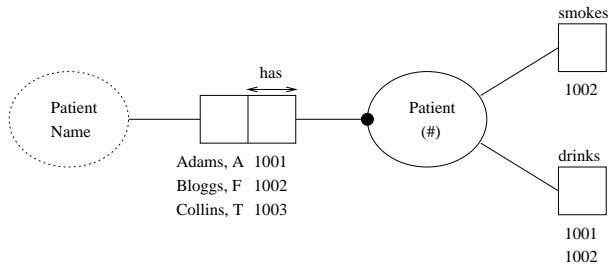
\EpsfFig[\HospitalA]{One way of modelling the hospital universe of discourse}

A third way to model this is generalise the smoking and drinking
fact types into a single binary, introducing the object type \SF{Vice}
(\SF{S = Smoking}, \SF{D = Drinking}) to maintain the distinction 
(see \SRef{\HospitalB}). 
Intuitively, most people would consider the schemas of \SRef{\HospitalA}
and \ref{\HospitalB} to be equivalent. 
Formally, this intuition can be backed by introducing \SF{Vice} as a
derived type to the schema of \SRef{\HospitalB}, and by specifying
exactly how the fact types of each can be translated into the fact types
of the other. 
For example, facts expressed in the first model may be expressed in
terms of the second model using the translations:
\begin{eqnarray*}
   \SF{Patient $p$ smokes} & \SF{IFF} & \SF{Patient $p$ indulges in Vice `S'}\\
   \SF{Patient $p$ drinks} & \SF{IFF} & \SF{Patient $p$ indulges in Vice `D'}
\end{eqnarray*}
and facts in the second model may be expressed in the first using the 
translation:
\begin{eqnarray*}
   \SF{Patient $p$ indulges in Vice $v$} 
   & \SF{IFF} & (\SF{Patient $p$ smokes AND Vice $v$ has ViceCode `S'})\\
   & \SF{OR}  & (\SF{Patient $p$ drinks AND Vice $v$ has ViceCode `D'})
\end{eqnarray*}
Even though the second schema might sometimes be considered more "natural",
the first schema will usually be more efficient to implement with respect 
to the number of relational tables.
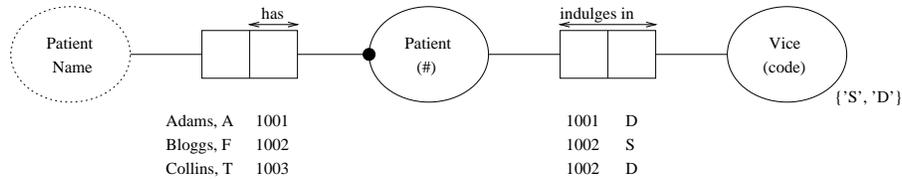
\EpsfFig[\HospitalB]{Another way of modeling the hospital universe of discourse}

\subsection{Predicate specialisation and generalisation}
In this subsection we consider a class of schema transformations known
as predicate specialisation, as well as its inverse, predicate
generalisation.
If two or more fact types may be thought of as special cases of a more
general fact type then we may replace them by the more general
fact type, as long as the original distinction can be preserved in 
some way. 
For example, if we transform the schema of \SRef{\HospitalA} into that
of \SRef{\HospitalB}, we generalise \SF{smoking} and \SF{drinking} into 
\SF{indulging in a vice}, where vice has two specific cases. 
If we transform in the opposite direction, we specialise \SF{indulging 
in a vice} into two fact types, one for each case.

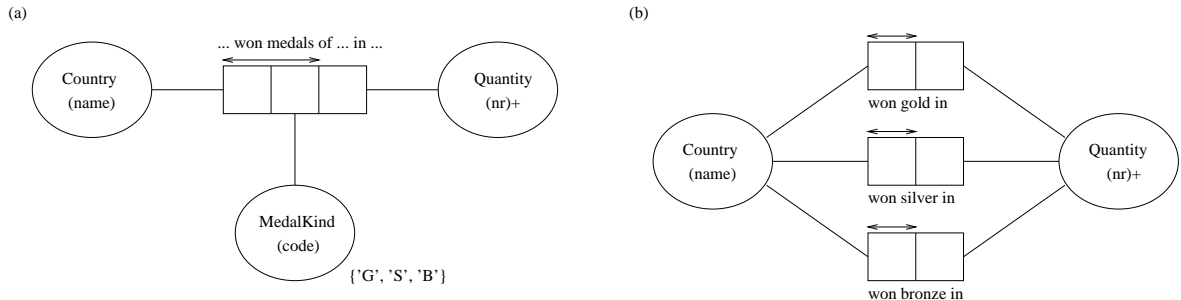
\EpsfFig[\Olympics]{Olympic Games universe of discourse}
A fact type may be specialised if a value constraint or a frequency 
constraint indicates that it has a finite number of cases. 
Examples with value constraints are more common, so we examine these first. 
The drinker-smoker example provides one illustration, where \SF{Vice} 
has the value constraint \SF{\{\SF{'S'},\SF{'D'}\}}. 
As another example, consider the Olympic Games schema depicted in
\SRef{\Olympics} (a). 
Because there are exactly three kinds of medal, the ternary may be
specialised into three binaries, one for each medal kind, as shown 
in \SRef{\Olympics} (b).
 
You may visualise the transformation from schema (a) into schema (b) 
thus: 
when the object type \SF{MedalKind} is absorbed into the ternary 
fact type, it breaks it up (or specialises it) into the three binaries.
The reverse transformation from (b) to (a) generalises the three
binaries into the ternary by extracting the object type \SF{MedalKind}.

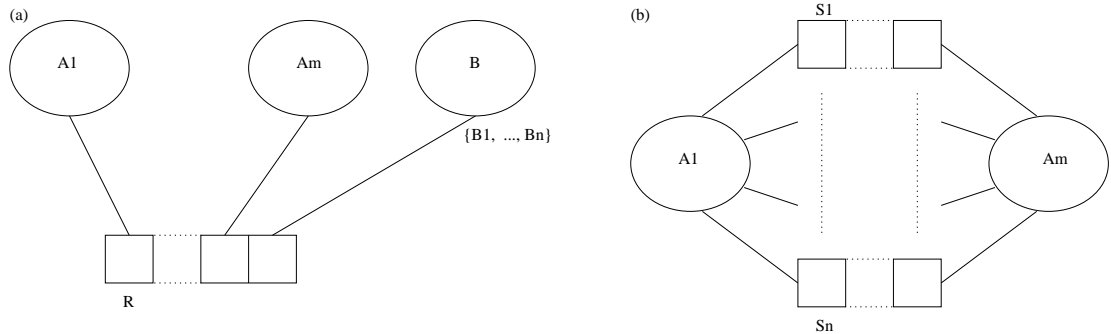
\EpsfFig[\OTAbsorb]{Object type absorption transformation}
Notice that in the vices example, a binary is specialised into unaries.
With the games example, a ternary is specialised into binaries. 
In general, when an $n$-valued object type is absorbed into a fact type,
the $n$ specialised fact types that result each have one less role than
the original (since the object type has been absorbed). 
This general result is set out in \SRef{\OTAbsorb}.
In this case, the answer to the question which alternative will be
most efficient depends on such things as the access and data profile
of the application in question.
Since we consider these transformations in terms of data schemas
in a conceptual modelling technique, we can actually involve end
users (domain experts) to give an indication of the access and data profiles.
This will generally be harder when only a database schema is available.

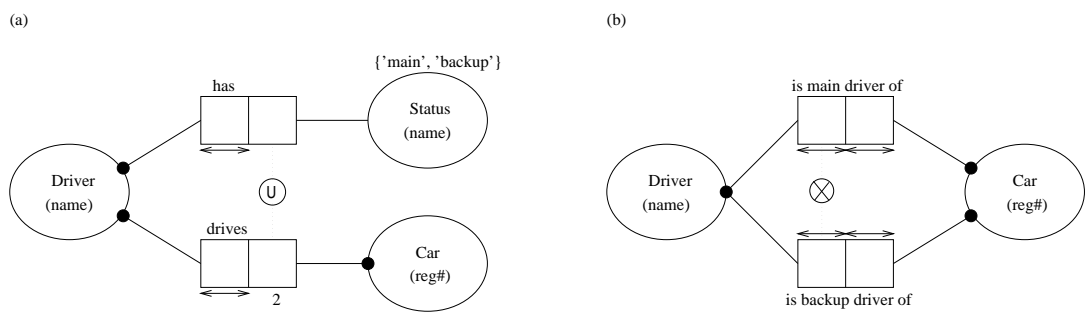
\EpsfFig[\Rally]{The drives fact type is specialised by absorbing Status}
As an introduction to a second schema transformation, consider 
\SRef{\Rally}. 
The two schemas provide alternative models for a fragment of a car 
rally application. 
The circled "u" is an external uniqueness constraint 
(each Status-Car combination applies to at most one driver) 
and the "2" is a frequency constraint.
Each car in the rally has two drivers (a main driver and a backup 
driver), and each person drives exactly one car. 
Schema (a) is transformed into schema (b) by absorbing the object type 
\SF{Status} into the drives fact type, specialising this into the main 
driver and backup driver fact types. 
The circled "X" is an exclusion constraint (no driver is both a main
driver and a backup driver).
The reverse transformation generalises the specific driver fact types 
into the general one by extracting the object type \SF{Status}. 
Since this object type appears in a different fact type, this 
equivalence does not fit the pattern of the transformation provided 
in \SRef{\OTAbsorb}.

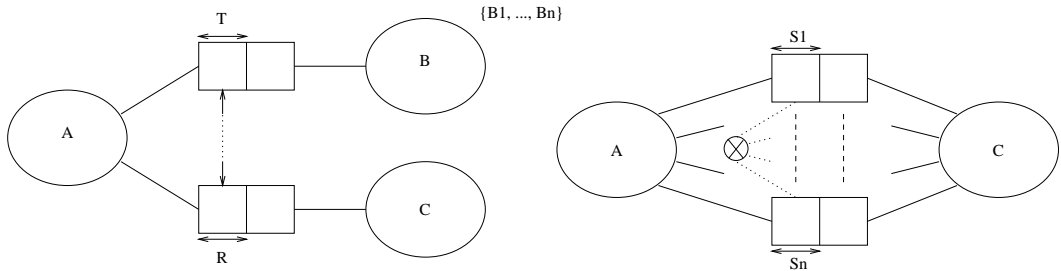
\EpsfFig[\Absorb]{Another absorption transformation}
Note how the constraints are transformed. 
The external uniqueness constraint in (a) says that each car has at most 
one main driver and at most one backup driver. 
This is captured in (b) by the uniqueness constraints on the roles of 
\SF{Car}. 
The uniqueness constraint on the drives fact type in (a) corresponds in 
(b) to the uniqueness constraints on the roles of \SF{Driver}. 
The uniqueness constraint on the status fact type in (a) is captured by 
the exclusion constraint in (b). 
The mandatory and frequency constraints on \SF{Car}'s role in (a) require the 
two mandatory role constraints on \SF{Car} in (b). 
Finally, the mandatory role constraints on \SF{Driver} in (a) are catered for 
in (b) by the disjunctive mandatory role constraint (shown explicitly 
here).
This example illustrates our second specialisation/generalisation 
transformation as shown in \SRef{\Absorb}.
In our example, \SF{A}, \SF{B} and \SF{C} correspond to \SF{Driver}, 
\SF{Status} and \SF{Car}; and the equality constraint is implied by the
two mandatory role constraints on \SF{Driver}. 

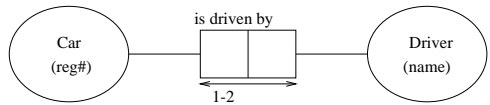
\EpsfFig[\CanSpec]{Can the predicate be specialised?}
Sometimes we may wish to transform a schema into another that is not 
quite equivalent. 
The reason for doing this may be that by slightly changing the
semantics of the schema, a much more efficient storage becomes
possible.
For example, suppose that in our car rally application we limit each car 
to at most two drivers, but do not classify the drivers in any 
meaningful way (e.g. as main or backup drivers).
Let us also remove the constraint that drivers may drive only one car. 
This situation is schematised in \SRef{\CanSpec}.

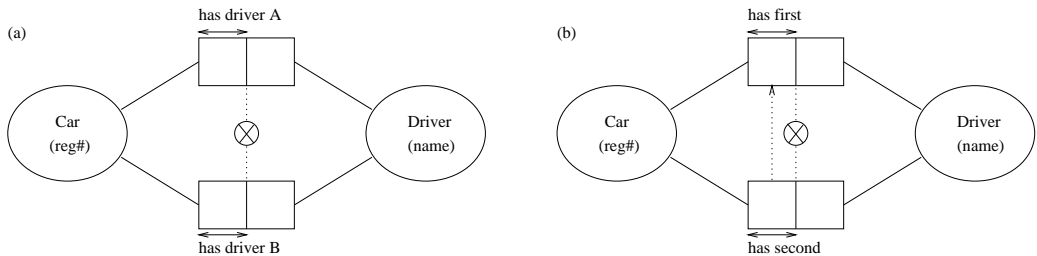
\EpsfFig[\Strengthen]{Two ways of strengthening}
Although no \SF{Status} object type is present, the frequency constraint in
Figure \ref{\CanSpec} tells us that each car has at most two drivers. 
This enables us to introduce an artificial distinction to specialise 
the fact type into two cases, as shown in \SRef{\Strengthen}. 
Since this distinction is not present in the original schema, the 
alternatives shown in \SRef{\Strengthen} are not equivalent to the original; 
they are in fact stronger. 
Schema (b) is actually stronger than schema (a). 
If the \SF{Car} role in \SRef{\CanSpec} is mandatory, then the Car roles in 
(a) are disjunctively mandatory, and the top \SF{Car} role in (b) is mandatory 
(which then implies the subset constraint (shown as a dotted arrow)).
In an application where other facts are stored about cars but not about 
drivers, one of these alternatives may well be chosen to avoid a separate 
table being generated for car-driver facts when the schema is mapped (the 
specialised fact types are functional rather than m:n). 
In practice, schema (b) of \SRef{\Strengthen} would normally be chosen.
Transforming from the original schema to one of those in \SRef{\Strengthen}
strengthens the schema by adding information. 
Transforming in the opposite direction weakens the schema by losing 
information. 
Any such transformations which add or lose information 
should be the result of conscious decisions which are acceptable to the 
client (for which the application is being modeled).
The rationale behind such a transformation is that although the number
of allowed populations might decrease, the resulting schema might be
much more efficient to implement.
If the excluded populations will actually never have to be stored, than
it certainly makes sense to make such a transformation.

This example illustrates our third specialisation/generalisation 
transformation, as shown in \SRef{\StrongerSch}. 
In practice, the transformation is usually performed from right to left 
(strengthening the schema), with a subset constraint added from the 
first role of \SF{S2} to that of \SF{S1} if $n = 2$. 
In cases like this, end user  participation in the 
optimisation process is almost essential as only the end users can really 
decide whether the strengthening of a schema, and thus limitation of
the underlying grammar, outweighs the expected efficiency gains.
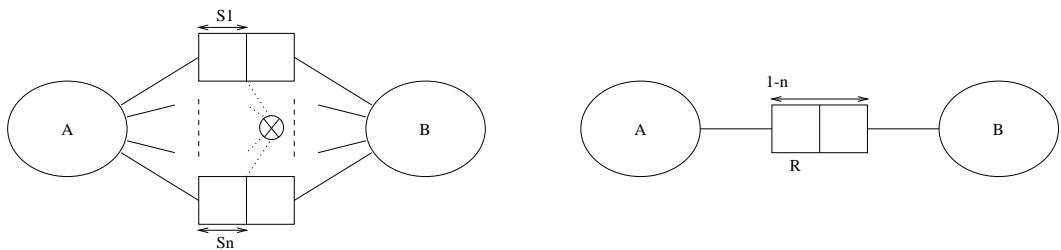
\EpsfFig[\StrongerSch]{The left-hand schema is stronger than the right-hand schema}

This completes our discussion of a small potpourri of schema transformations.
As stated before, a wider range of schema transformations can be found
in \cite{PhdThesis:89:Halpin:NIAMForm} and \cite{Book:94:Halpin:ORM}.

   \section{Conceptual Schema Universe}
\SLabel{section}{ORMU}

Even though formalisations of ORM have been published before 
(\cite{PhdThesis:89:Halpin:NIAMForm},
 \cite{Report:91:Hofstede:PSM}, \cite{Report:91:Hofstede:LISA-D},
 \cite{Report:94:Brouwer:ORMKernel}, \cite{Report:94:Halpin:ORMPoly}), 
we provide such a formalisation once more to be self contained.
However, in this formalisation we limit ourselves to syntactical
issues only. 
Issues regarding semantics can be found in the referenced publications.
Furthermore, a related formalisation is provided in 
\cite{Report:95:Proper:CDMKernel}.
The formalisation given there is based on a smaller number of
basic concepts while maintaining the full expressibility of ORM.

We also introduce the concept of a universe of ORM models which is
to be used as a medium for schema transformations.
The next section then discusses what a schema version is within this
ORM universe.
The notion of having a universe of data schemas, and the data schema of
a universe of discourse describing a journey (evolution) through this 
universe as a sequence of versions has been 
introduced before in the field of evolving information systems 
(\cite{Report:92:Proper:EvolvAM},
 \cite{Report:93:Proper:EvolvPSM}, 
 \cite{PhdThesis:94:Proper:EvolvConcModels}).

\subsection{Information structure universe}
We assume the reader has a basic working knowledge of the concepts 
underlying ORM or ER.
A conceptual schema is presumed to consist of a set of types $\Types$.
This set can be divided in three subclasses.
The first class is the set of object types (entity types) $\ObjTypes$.
Within this class a subclass of value types 
$\ValueTypes \subseteq \ObjTypes$ can be distinguished.
Instances from value types usually originate from some underlying domain 
such as strings, natural numbers, audio, video, etc.
Later a function is introduced that assigns a domain (set of values)
to each value type.
A separate class of types are the relationship types $\RelTypes$.
We now have the following type classes:
\begin{eqnarray*}
   \Types      & \Eq       & \ObjTypes \union \RelTypes\\
   \ValueTypes & \subseteq & \ObjTypes
\end{eqnarray*}

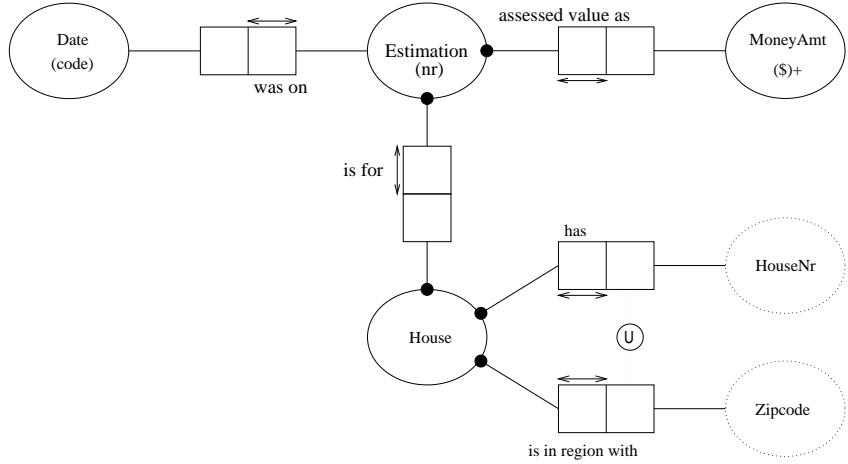
\EpsfFig[\ExampleIS]{Example information structure}
An example ORM conceptual schema can be found in \SRef{\ExampleIS}. 
For this schema we have:
\begin{eqnarray*}
   \ValueTypes_i  &\Eq& \setje{ \SF{DateCode}, \SF{\$Value}, \SF{HouseNr}, 
                                \SF{Zipcode}, \SF{EstimationNr} }\\
   \ObjTypes_i    &\Eq& \setje{ \SF{Date}, \SF{MoneyAmt}, \SF{House}, 
                                \SF{Estimation}, \SF{DateCode}, 
                                \SF{\$Value}, \SF{HouseNr}, 
                                \SF{Zipcode}, \SF{EstimationNr}  }\\
   \RelTypes_i    &\Eq& \setje{ \SF{was on}, \SF{assessed value as}, \SF{is for}, 
                                \SF{has}, \SF{is in region with} }
\end{eqnarray*}
where the index $i$ indicates that we are dealing with schema version $i$.

Types are interrelated in a number of different ways. 
In the ORM universe, we have the following relationships between types:
\begin{enumerate}
  \item Relationship types consist of a number of {\em roles} (also
        referred to as {\em predicators}). The roles of an ORM schema
        are captured in the set $\Preds$. The roles in $\Preds$ are
        distributed among the relationship types by the partition:
            $\Roles: \RelTypes \Func \Powerset^+(\Preds)$. 
        (Note that $\Powerset^+(\Preds)$ yields all non-empty subsets
         of $\Preds$). 
        The object types playing each of the roles are yielded by the function
        $\Player: \Preds \Func \Types$. 

        For the $\Roles$ function, we have the following `inverse'
        function returning the relationship to which a given role
        belongs: 
           $\Rel: \Preds \Func \RelTypes$, 
        which is defined by: $\Rel(r) = f  \iff r \in \Roles(f)$.

  \item The inheritance of properties is captured by the 
        $\SubOf \subseteq \ObjTypes \Carth \ObjTypes$ relation; 
        with the intuition:
        \begin{quote}
           if $x \SubOf y$ then `the population of $x$ is a subset of 
           the population of $y$'
        \end{quote}
        One of the reductions in the number of basic concepts for
        ORM models as reported in 
        \cite{Report:95:Proper:CDMKernel} is the integration of 
        traditional subtyping and 
        polymorphism (\cite{Report:94:Halpin:ORMPoly}), also known as 
        categorisation in EER (\cite{Article:85:Elmasri:ERSetModel}).
        This is the $\SubOf$ relationship, which is strictly concerned
        with inheritance of populations between type.
        For a more detailed discussion on the relationship to
        traditional subtyping and polymorphism, refer to
        \cite{Report:95:Proper:CDMKernel}.

  \item An information structure version within the universe has to
        adhere to certain well-formedness rules.
        An information structure version is fully determined by the set of
        types contained in it.
        Therefore we presume to have the following well-formedness
        predicate: $\IsIS \subseteq \Powerset(\Types)$.
        The exact definition of this predicate follows in the next section.
\end{enumerate}
An information structure universe $\InfStructUniverse$ over a set of domains 
$\Doms$ is determined by the components of the following tuple:
\[ \tuple{
     \RelTypes, \ObjTypes, \Preds, \SubOf, \Roles, \Player, \IsIS 
} \]
The first 3 components provide the types and roles present in the 
information structure, and the last 4 components describe their mutual 
relationships; providing the `fabric' of the information structure. 

Relationship types and object types are exclusive:
\begin{Axiom}{ISU}(type exclusion)
   $\RelTypes \intersect \ObjTypes = \emptyset$
\end{Axiom}
The careful reader may now raise the question: {\em whatever happened
to objectification/nesting of relationship types?}
This question will be answered at the end of this section.

For the relationships between types some well-formedness rules apply. 
Firstly, the inheritance hierarchy is both transitive and irreflexive:
\begin{Axiom}{ISU}(transitive)  
   $x \SubOf y \SubOf z \implies x \SubOf z$
\end{Axiom}
\begin{Axiom}{ISU}(irreflexive) 
   $\lnot x \SubOf x$
\end{Axiom}
The separation between value types and non value types must be maintained
in the inheritance hierarchy:
\begin{Axiom}{ISU}(separation)
   If  $x \SubOf y$, then: 
   \( x \in \ValueTypes \iff y \in \ValueTypes \).
\end{Axiom}
From the transitive $\SubOf$ relation we can derive the intransitive one ($\SubOf_1$) as follows:
\[ x \SubOf_1 y \iff x \SubOf y \land \lnot\Ex{z}{x \SubOf z \SubOf y} \]
The finite depth of the identification hierarchy in ORM is expressed by the following schema of 
induction:
\begin{Axiom}{ISU}(identification induction) \Eol
   If $F$ is a property for object types, such that:
      \[ \mbox{for any $y$, we have:~}  \Al{x:y \SubOf_1 x}{F(x)} \implies F(y) \]
   then $\Al{x \in \ObjTypes}{F(x)}$
\end{Axiom}
The latter axiom was not explicitly present in the previous discussions, 
but was always presumed to be implicitly present. 
In this paper, it is only stated for reasons of completeness. 
Note that the identification induction schema can be proven from the properties
of $\SubOf$ if the axiomatic setup were extended with the natural numbers and 
their axioms (in particular natural number induction). 
In our formalisation, however, we do not presume the presence of the natural 
numbers.

\subsection{Conceptual schema universe}
A conceptual schema universe $\CSUniverse$, over a set of concrete domains
$\Doms$, for ORM is now identified by the following components:
\[  \tuple{\InfStructUniverse,\Constraints,\DerRules,\Dom,\IsSch} \]
These components are:
\begin{enumerate}
   \item Each conceptual schema contains an information structure as its
         backbone. 
         So the information structure universe is part of the schema universe.

   \item The first set of rules are the constraints. 
         They must be taken from the set of constraints $\Constraints$.
         As a textual language to define constraints one may choose e.g. 
         FORML or LISA-D.
         ORM also has a graphical representation for the most generally used 
         constraints such as uniqueness, mandatory (totality) and 
         exclusiveness constraints.

   \item As some of the types in the information structure are 
         derivable from other types, conceptual schemas may contain 
         derivation rules.
         The set of derivation rules is provided as $\DerRules$.

         In this article we do not elaborate on a language for the definition
         of these rules.
         However, we do presume the existence of two functions on
         derivation rules:
         \[ \Defines: \DerRules \Func \Types \]
         yielding the relationship type that is being defined by a 
         derivation rule, and
         \[ \Depends: \DerRules \Func \Powerset(\Types) \]
         returning the types to which the derivation rule (directly) 
         refers.

   \item All atomic value types have associated a domain of 
         pre-defined denotable values. 
         For instance, the value type \SF{Nat no} will typically
         have associated some subset of the natural number.
         Formally, the possible relationships between the atomic value
         types and the domains is provided as:
         $\Dom = (\AtomTypes \intersect \ValueTypes) \Func \Doms$,
         where $\Doms$ denotes the set of domains.

   \item Finally, not all schemas in the universe of schemas spanned by
         the above components correspond to a proper schema.
         Therefore the predicate $\IsSch$ is introduced to distinguish the
         proper schema versions. 
         Its formal definition will follow below.
\end{enumerate}

\subsection{Objectified relationship types and other complex types}
The first axiom we formulated on type classes was the
fact that the set of relationship types is disjoint from the set of 
object types.
This requirement seems to be in contradiction with the existence
of so-called objectified relationship types.
In ORM, as well as ER, one can objectify relationship types and
as of then treat them as if they were object types.
An example of this is shown in the left hand side of \SRef{\Objectif}.
An \SF{Enrollment} is both used as a relationship type between
\SF{Student} and \SF{Subject} and as an object type playing a role
in the \SF{resulted in} relationship type.
It is a well known fact that semantically objectifications are equivalent
to co-referenced object types 
(i.e. object types that require a combination of two or more
reference types to identify them).
In the right hand side of \SRef{\Objectif}, the equivalent schema using
a co-referenced object type \SF{Enrollment} is shown.
\EpsfFig[\Objectif]{Example objectification}

In this article we will treat all objectifications as if they were 
co-referenced object types.
The reason for doing this is that it reduces the complexity of the 
formalisation; it leads to a reduction of the number of type classes.
Nevertheless, when modelling a universe of discourse one would
still like to be be able to use objectification simply because
it sometimes leads to a more natural way of modelling domains (see \SRef{ConcEq}).
For example, for some given universe of discourse, the top schema 
of \SRef{\Objectif} may be much more natural than the bottom schema.
Although the two depicted schemas are mathematically equivalent, they
are not likely to be equally preferable.

Objectification may be seen as a form of abstraction.
In terms of the example, the \SF{Enrollment} objectification is an 
abstraction of the underlying relationship types \SF{is in} and 
\SF{for}.
Objectification, and abstraction in general, can therefore be seen
as a third dimension for a flat conceptual model.
The flat model consists of relationship types and object types 
as defined for the information structure universe, while the
third dimension is concerned with abstractions, like objectification.
This third dimension is concerned only with abstraction, and therefore
has no influence on mathematical equivalence.
Therefore, for schema transformations aimed at optimisation of data schemas
one may ignore abstraction mechanisms like objectifications.

In \cite{Article:94:Campbell:AbstractionORM},
\cite{Report:95:Proper:CDMKernel} and 
\cite{Report:95:Campbell:Abstraction} a more elaborate discussion
is provided on abstractions as a third dimension for flat
conceptual modelling.
Finally, the above discussion for objectification also holds for
other complex types like set types, bag types and sequence types.
On the left hand side of \SRef{\Convoy} an example of a set
type is shown.
A \SF{Convoy} consists of a set of ships, where each ship is
identified by a unique code.
Convoys are not identified by some surrogate code but by the
actual set of ships in the convoy.
In the right hand side of this figure the flat view, without
abstraction, is shown.
The circled \SF{eu} signifies an {\em existential
uniqueness} constraint (\cite{Report:94:Hofstede:EU},
\cite{Report:94:Hofstede:EU-Identity}).
The existential uniqueness constraint in this case enforces the
rule that no two different convoys have the same set of ships
associated.
In this example, \SF{Convoy} is a set type and can be seen as
an abstraction of the underlying \SF{consisting of} relationship
type.
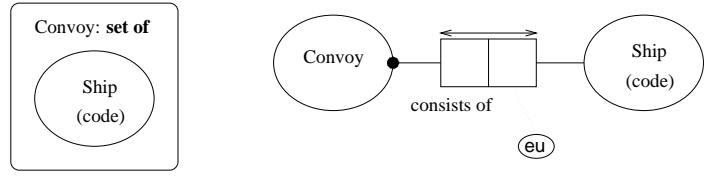
\EpsfFig[\Convoy]{The convoy set type example}

   \section{Versioning within a Universe}
\SLabel{section}{ORMV}

Once the user community has agreed that a certain conceptual schema is the 
preferred representation of the universe of discourse, that schema should
remain fixed during the optimisation process 
(unless a new optimisation is displayed to and then preferred by the user).
The schema may be implemented differently, but the approved 
conceptual schema should remain as is from a user's point of view.
This means that the types of the original schema should not be removed
during schema optimisations.
This is why we will make a distinction between so called {\em
conceptual types} and {\em internal types}.
The types of the original conceptual schema are obviously all marked as 
conceptual types.
Any types introduced for schema optimisation purposes are marked as
internal, and should not be 'accessible' directly by users.

When a conceptual type is transformed to a (set of) internal type(s),
a derivation rule is specified that derives the population of the 
transformed type.
This means that the user can still access the information in the database
in terms of the conceptual schema.
For example using a conceptual query language like LISA-D
(\cite{Report:91:Hofstede:LISA-D}).
Conversely, when updating the database, the user would like to do this
in terms of the conceptual schema as well.
This means that for the transformed conceptual types, update rules must be
specified which translate updates of the conceptual types to updates of
the proper internal types.
In a later section we provide extra rules to enforce certain well-formedness
rules on the derivation rules in combination with the update rules
(e.g. to avoid view update problems).

In the remainder of this section we discuss what constitutes a schema
version within a schema universe.
To identify a schema version $\Schema_i$, the following components are 
required:
\begin{enumerate}
   \item A set of types $\Types_i \subseteq \Types$ which identifies the 
         information structure version.
   \item The internal types are types: $\ITypes_i \subseteq \Types_i$.
   \item A set of constraints $\Constraints_i \subseteq \Constraints$ that 
         should hold for this version.
   \item A set of derivation rules $\DerRules_i \subseteq \DerRules$ for the 
         derivable types in this version.
   \item A set of update rules $\UpdRules_i \subseteq \DerRules$.
   \item A domain mapping $\Dom_i \in \Dom$ for the value types of this version.
\end{enumerate}
The set of types $\Types_i$ spans the information structure version
$\InfStruct_i$ for this conceptual schema version.
Sometimes we have to refer to a specific type class of schema components
within one information structure version.
These classes are derived as follows:
\begin{eqnarray*}
   \ObjTypes_i   & \Eq & \Types_i \intersect \ObjTypes\\
   \ValueTypes_i & \Eq & \Types_i \intersect \ValueTypes\\
   \RelTypes_i   & \Eq & \Types_i \intersect \RelTypes\\
   \Preds_i      & \Eq & \Union\Sub{r \in \RelTypes_i} \Roles(r)
\end{eqnarray*}

\subsection{Well-formedness of information structure versions}
We now first focus on well-formedness of information structure versions.
These rules mainly deal with completeness of a single version.

If a role is part of the current version, then so must the player of the 
role:
\begin{Axiom}{ISV}
   $p \in \Preds_i \implies \Player(p) \in \Types_i$
\end{Axiom}
Types in this version that are lower in the type hierarchy ($\SubOf$)
must have at least one forefather present:
\begin{Axiom}{ISV} 
   $x \in \ObjTypes_i \land x \SubOf z \implies 
       \Ex{y \in \Types_i}{x \SubOf y}$ 
\end{Axiom}
A schema version should be a connected graph:
\begin{Axiom}{ISV}
   The graph $\tuple{\Types_i,E_i}$ where
   \[ E_i = \Set{\tuple{\Player(p),\Rel(p)}}{p \in \Preds_i} \union
            \Set{\tuple{x,y}}{x \IdfBy y} \]
   should be connected.
\end{Axiom}
Together the \ClassRef{ISV} axioms define the $\IsIS$ predicate on
sets of types:
\begin{definition}
   Let $X \subseteq \Types$, then
   \[ \IsIS(X) ~\Eq~ \mbox{the information structure version spanned by $X$
                           adheres to the \ClassRef{ISV} axioms} \]
\end{definition}

\subsection{Conceptual schema versions}
For schema versions well-formedness rules can be formulated as well.
In this article, we do not discuss all rules that could be formulated, but 
limit ourselves to the bare essentials.
For example, in previous formalisations one can find rules requiring
the existance of proper identification schemes (reference schemes)
that provide for all types a proper denotation of their instances
in terms of values.
An identification scheme is a property that is relevant for a
data schema that is a conceptual schema, and does therefore not play
a role during data schema optimisations.

The following rules are relevant for schema versions during an
optimisation process.
Domain assignment for value types within a version must be complete:
\begin{Axiom}{CSV}(complete domain assignment)
   $\Dom_i: (\ValueTypes_i \intersect \BasicTypes) \Func \Doms$
\end{Axiom}
Only one derivation (and update) rule can be defined for
a type:
\begin{Axiom}{CSV}(unique rules)
   If $r,s \in \UpdRules_i$ or $r,s \in \DerRules_i$, then:
   \[ \Defines(r) = \Defines(s) \implies r = s \]
\end{Axiom}
Finally, an update rule must be specified for all the
internal types which are populatable:
\begin{Axiom}{CSV}(update rule completeness)
   \( \Set{\Defines(r)}{r \in \UpdRules_i} = \ITypes_i \)
\end{Axiom}
As stated before, more rules could be added but have been omitted.
For a more elaborate discussion of well-formedness of ORM model
versions, refer to \cite{Report:91:Hofstede:PSM}, 
\cite{Report:91:Hofstede:PSMPromo},
\cite{Report:94:Brouwer:ORMKernel}, or \cite{Report:94:Halpin:ORMPoly}.
In actual fact, some of the extra rules that could be added are
based on which particular ORM version (or ER version for that matter)
one uses for conceptual modelling.

We are now finally in a position to define exactly what a proper ORM schema is:
\begin{definition}
   Let 
     $\Schema_i \Eq 
       \tuple{\Types_i,\ITypes_i,\Constraints_i,\DerRules_i,\UpdRules_i,\Dom_i}$
   such that:
   \[
      \Types_i \subseteq \Types,~ 
      \ITypes_i \subseteq \Types_i,~
      \Constraints_i \subseteq \Constraints,~
      \DerRules_i \subseteq \DerRules,~
      \UpdRules_i \subseteq \UpdRules,~\mbox{\it and}~
      \Dom_i: \ValueTypes \PartFunc \Doms
   \]
   then:
   \[ \IsSch(\Schema_i) ~\Eq~ 
      \mbox{$\Schema_i$ adheres to the \ClassRef{CSV} axioms}
      \land \IsIS(\Types_i) \]
\end{definition}
The set of schemas in a conceptual schema universe is now defined as:
\[ \Schema = \Powerset(\Types) \Carth \Powerset(\Types) \Carth 
             \Powerset(\Constraints) \Carth 
             \Powerset(\DerRules) \Carth \Powerset(\DerRules) \Carth
             \Dom
\]
Note that this carrier set contains both correct and incorrect schemas.

   \section{Formalisation of Transformations}
\SLabel{section}{TransLang}

In this section we discuss a language to define the actual 
schema transformations.
A general format is introduced in which classes of schema transformations 
can be defined using a so called {\em data schema transformation scheme}. 
When applying such a transformation scheme to a concrete data schema,
the transformation scheme is interpreted in the context of that
schema, leading to a concrete transformation.

We do not provide a fully elaborated and formalised language to 
specify transformation schemes.
Doing so would take up too much space and is not expected to lead to 
a better understanding of the problems addressed in this article. 
We do, however, discuss an example of a transformation scheme and
an application to an existing data schema.

\subsection{A language for transformation schemes}
A schema transformation scheme is typically centered around a set of
parameters.
When applying such a scheme on an actual data schema, these
parameters have to be instantiated with actual components from the 
data schema that is to be transformed.

\begin{table}[htb]
\[ \begin{array}{l}
   \SF{Transformation schema} \mbox{\it ~OTEmission~} 
      ( x!n, (r!n)!m, s!n, t, y, v, w, l, d, i!m);\\
   \Tab \SF{Object types:}\\
   \Tab \Tab x!n, y, l;\\
   \\
   \Tab \SF{Value types:}\\
   \Tab \Tab l:d;\\
   \\
   \Tab \SF{Relationship types:}\\
   \Tab \Tab (f = [ (x:r)!n ])!m,\Eol
   \Tab \Tab g = [ (x:s)!n, y:u ],\Eol
   \Tab \Tab h = [ y:v, l:w ];\\
   \\
   \Tab \SF{Constraints:}\\
   \Tab \Tab \SF{c1: UNIQUE}~ \setje{v};\Eol
   \Tab \Tab \SF{c2: UNIQUE}~ \setje{w};\Eol
   \Tab \Tab \SF{c3: MANDATORY}~ \setje{v};\Eol
   \Tab \Tab \SF{c4: EACH $l$ IS IN $i!m$};\\
   \\
   \Tab \SF{From:}\\
   \Tab \Tab f!m;\\
   \\
   \Tab \SF{To:}\\
   \Tab \Tab y, l, d, g, h, \SF{c1}, \SF{c2}, \SF{c3}, \SF{c4};\\
   \\
   \Tab \SF{Derivation rules:}\\
   \Tab \Tab (f = \SF{PROJ}[(r=s)!n]~
                  \SF{SEL}[u = v, w = i]~
                  g~\SF{JOIN}~h)!m;\\
   \\
   \Tab \SF{Update rules:}\\ 
   \Tab \Tab h = \SF{UNION OF} (\SF{PROJ}[(s=r)!n, u=\SF{Val}(y,i)] f)!m;\Eol
   \Tab \Tab g = \setje{\tuple{v=\SF{Val}(y,i),w=i}!m}\\
   \\
   \SF{End Transformation schema.}
\end{array} \]
\SLabel{table}{ExTransScheme}
\caption{An example transformation scheme}
\end{table}

An example transformation schema is given in \SRef{ExTransScheme}.
It is the textual denotation of the schema transformation
scheme underlying the example transformation depicted in \SRef{\OTAbsorb}
(going from (b) to (a)).
The parameters to this transformation schema are
$x!n, (r!n)!m, s!n, t, y, v, w, l, d, i!m$. 
This is a set of parameters with a variable size. 
The expression $x!n$ denotes the list of parameters
$x_1,\ldots,x_n$, where $n$ is not a-priori known.
Analogously, $(r!n)!m$ denotes the sequence of parameters:
\[ r_{1,1},\ldots,r_{n,1},\ldots,\ldots,r_{1,m},\ldots,r_{n,m} \]
So this transformation scheme takes $n + n m + n + 4 = n(m+2)+4$ parameters,
where $n$ and $m$ are to be determined at the time of application.
In \SRef{\TransforEx} we have depicted a graphical representation
(omitting derivation and update rules) of this transformation scheme.
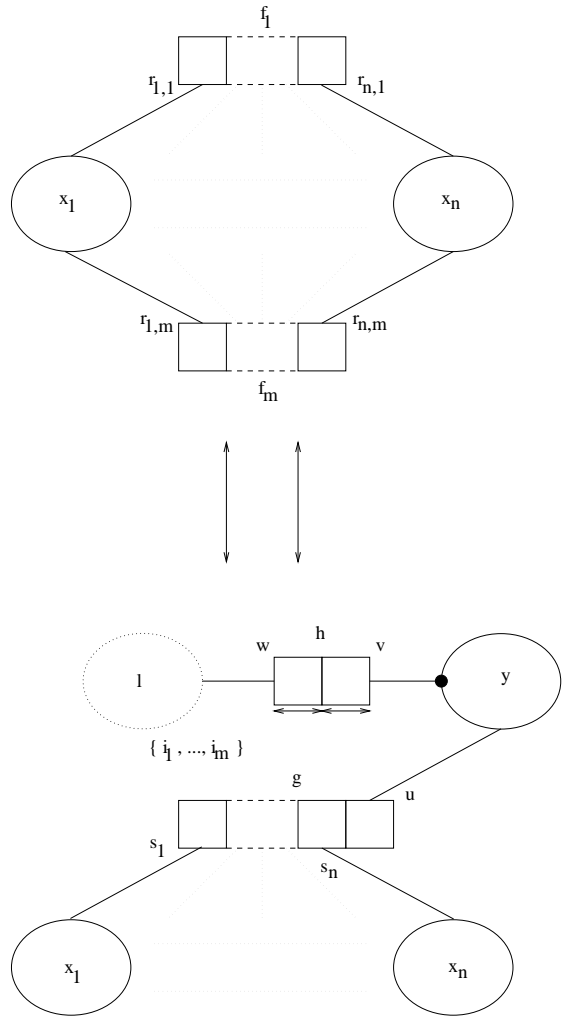
\EpsfFig[\TransforEx]{Graphical representation of transformation scheme}

Please note that relationship $h$ between object type $y$ and
value type $l$ provides the identification for object type $y$.
This relationship is normally implicitly represented by placing $(l)$
inside the ellipse for object type $y$.
For example, in \SRef{\Olympics}, the \SF{MedalKind} object type
is identified via the value type \SF{code}. 
The relationship between this object type and the value type is not
drawn explicitly.

A concrete example of an application of the above transformation
scheme would be:
\[ \arraycolsep=0.2em \small \SmallSize \begin{array}{ll}
      {\it OTEmission}( &[\SF{Country},\SF{Quantity}],\Eol
                        & [~[\SF{won-gold-in-1},\SF{won-gold-in-2}],
                            [\SF{won-silver-in-1},\SF{won-silver-in-2}],
                            [\SF{won-bronze-in-1},\SF{won-bronze-in-2}]~],\Eol
                        & [\SF{won-medals-of-in-1},\SF{won-medals-of-in-3}],\Eol
                        & \SF{won-medals-of-in-2}, \SF{MedalKind}, 
                          \SF{MedalKind.code-1}, \SF{MedalKind.code-2}, 
                          \SF{char},\Eol
                        & [\SF{'G'},\SF{'S'},\SF{'B'}]~~) 
\end{array} \]   
which is the textual representation of the transformation from (b) to (a)
in \SRef{\Olympics}.
Here we have used the convention that \SF{won-gold-in-1}, 
\SF{won-gold-in-2} refer to the first and second roles of the fact type 
labelled \SF{won gold in} respectively.
Furthermore, \SF{MedalKind.code-1} and \SF{MedalKind.code-2} refer to the
roles of the relationship type that is implicitly present between the
object type \SF{MedalKind} and the value type \SF{code}.
This relationship type in itself will be referred to by \SF{MedalKind.code}.

This leads to the following instantiation of the variables for the 
transformation scheme:
\[ \arraycolsep=2pt \begin{array}{lllllllll}
      x_1     & = & \SF{Country}            &~~
      x_2     & = & \SF{Quantity}           &~~
      r_{1,1} & = & \SF{won-gold-in-1}      \\
      r_{2,1} & = & \SF{won-gold-in-2}      &~~
      r_{1,2} & = & \SF{won-silver-in-1}    &~~
      r_{2,2} & = & \SF{won-silver-in-2}    \\
      r_{1,3} & = & \SF{won-bronze-in-1}    &~~
      r_{2,3} & = & \SF{won-bronze-in-2}    &~~
      s_1     & = & \SF{won-medals-of-in-1} \\
      s_2     & = & \SF{won-medals-of-in-3} &~~
      t       & = & \SF{won-medals-of-in-2} &~~
      y       & = & \SF{MedalKind}          \\
      u       & = & \SF{MedalKind.code-1}   &~~
      v       & = & \SF{MedalKind.code-2}   &~~
      l       & = & \SF{code}               \\
      d       & = & \SF{char}               &~~
      i_1     & = & \SF{'G'}                &~~
      i_2     & = & \SF{'S'}                \\
      i_3     & = & \SF{'B'}
\end{array} \] 
This allows us to further fill in the transformation scheme.
The \SF{Object types} statement simply requires that $x_1,\ldots,x_n,y,l$
are object types in the ORM universe. 
In our case we have to verify that \SF{Country}, \SF{Quantity},
\SF{MedalKind} and \SF{code}
are object types; which they indeed are.
By the \SF{Value types} statement it is required that $l$ be a value
type with underlying domain $d$.

A \SF{Relationship types} statement requires the presence of a proper
relationship type in the universe of ORM schemes.
In our case we have the requirements:
\begin{enumerate}
   \item for each $1 \leq i \leq m$
         there is a relationship type $f_i$ in the ORM universe such that
         we have
         $\Roles(f_i) = \setje{r_{1,i},\ldots,r_{n,i}}$ and furthermore
         $\Al{1 \leq j \leq n}{\Player(r_{j,i}) = x_j}$.
   \item there is a relationship type $g$ in the ORM universe such that
         $\Roles(g) = \setje{s_1,\ldots,s_n}$ and furthermore
         $\Al{1 \leq j \leq n}{\Player(s_j) = x_j}$.
   \item a relationship type $h$ exists in the ORM universe such that
         $\Roles(h) = \setje{v,w}$, $\Player(v) = y$ and $\Player(w) = l$.
\end{enumerate}
It is easy to verify that this holds for the running example with:
\[ f_1 = \SF{won-gold-in}, f_2 = \SF{won-silver-in}, 
   f_3 = \SF{won-bronze-in}, g = \SF{won-medals-of-in},
   h = \SF{MedalKind.code} \]

With the \SF{Constraints} statements we capture the requirement that
there exists a constraint $c4$ in the universe with definition
\SF{EACH $l$ IS IN $i!m$}. 
In our example case this becomes 
\[ \SF{EACH code IS IN 'G', 'S', 'B'} \]
The constraints $c1, c2$, and $c3$, are needed to ensure that 
value type $l$ can be used to properly identify the instances of
entity type $y$.
This is part of the graphical convention regarding the use of $(l)$,
in the example this is the relation between \SF{(code)} and \SF{MedalKind}.

What follows is a listing of components that are to be replaced
(\SF{From}) by the transformation.
In the running example these components are: \SF{won-gold-in}, 
\SF{won-silver-in}, and \SF{won-bronze-in}.
Similarly the \SF{To} statement lists the components added
by the transformation.
In the example they are:
\SF{MedalKind}, \SF{code}, \SF{char}, \SF{won-medals-of-in},
\SF{MedalKind.code} and constraints $c1$ to $c4$.

The \SF{Update rules} and \SF{Derivation rules} provide the translation of
populations between the schema before and after the transformation.
We have specified these rules in a textual representation of
a relational algebra like language.
As mentioned before, in this article we are not concerned with 
a concrete language for these purposes.
One statement in the update rules requiring some explanation though is
the $\SF{Val}(y,i)$ statement.
This function should generate a (unique) instance, for object type $y$,
which is identified by the value $i$.
Remember that instances of $y$ are identified by instances of value type $l$.
Value $i$ is an instance of $l$, and $\SF{Val}(y,i)$ simply
refers to the associated instance of $y$.
It is not legal to just take $\SF{Val}(y,i) = i$ since this would
lead to a situation where the population of $y$ and $l$ overlap,
which would be in contradiction with the fact that $y$ and $l$ are 
distinct object types that are not part of the same type hierarchy.
Actually, as $y$ is not a value type, $\SF{Val}(y,i)$ should lead to
a value that is not directly representable on a communication medium;
it should be an {\em abstract}\ instance.
So $\SF{Val}$ should provide some global encoding schema of value type
instances into abstract non-value type instances.
Finally, the completely substituted transformation is shown in 
\SRef{ExTransformation}.

\begin{table}[htbp]
\[ \SmallSize \SF{\begin{tabular}{l}
   Transformation schema \mbox{\it ~OTEmission}~;\\
   \Tab Object types:\\
   \Tab \Tab Country, Quantity, MedalKind, code;\\
   \Tab Value types:\\
   \Tab \Tab code: char;\\
   \Tab Relationship types:\\
   \Tab \Tab won-gold-in = [Country:won-gold-in-1, Quantity:won-gold-in-2];\\
   \Tab \Tab won-silver-in = [Country:won-silver-in-1, 
                              Quantity:won-silver-in-2];\\
   \Tab \Tab won-bronze-in = [Country:won-bronze-in-1, 
                              Quantity:won-bronze-in-2];\\
   \Tab \Tab won-medals-of-in = [Country:won-medals-of-in-1,
                                 Quantity:won-medals-of-in-3,
                                 MedalKind:won-medals-of-in-2];\\
   \Tab \Tab MedalKind.code =\\
   \Tab \Tab \Tab [MedalKind:MedalKind.code-1,code:MedalKind.code-2];\\
   \Tab Constraints:\\
   \Tab \Tab c1: UNIQUE \{ \SF{MedalKind.code-1} \};\\
   \Tab \Tab c2: UNIQUE \{ \SF{MedalKind.code-2} \};\\
   \Tab \Tab c3: MANDATORY \{ \SF{MedalKind.code-2} \};\\
   \Tab \Tab c4: EACH MedalKind IS IN 'G', 'S', 'B';\\
   \Tab From: won-gold-in, won-bronze-in\\
   \Tab To: MedalKind, code, char, MedalKind.code, won-medals-of-in, 
             c1, c2, c3, c4;\\
   \Tab Derivation rules:\\
   \Tab \Tab won-gold-in =\\
   \Tab \Tab \Tab PROJ[won-gold-in-1 = won-medals-of-in-1,
                            won-gold-in-2 = won-medals-of-in-3]\\
   \Tab \Tab \Tab \Tab SEL[won-medals-of-in-2 = MedalKind.code-1,
                           MedalKind.code-2 = 'G']~
                       won-medals-of-in JOIN MedalKind.code\\
   \Tab \Tab won-silver-in =\\
   \Tab \Tab \Tab PROJ[won-silver-in-1 = won-medals-of-in-1,
                            won-silver-in-2 = won-medals-of-in-3]\\
   \Tab \Tab \Tab \Tab SEL[won-medals-of-in-2 = MedalKind.code-1,
                           MedalKind.code-2 = 'S')]~
                       won-medals-of-in JOIN MedalKind.code\\
   \Tab \Tab won-bronze-in =\\
   \Tab \Tab \Tab PROJ[won-bronze-in-1 = won-medals-of-in-1,
                            won-bronze-in-2 = won-medals-of-in-3]\\
   \Tab \Tab \Tab \Tab SEL[won-medals-of-in-2 = MedalKind.code-1.
                           MedalKind.code-2 = 'B')]~
                       won-medals-of-in JOIN MedalKind.code\\
   \Tab Update rules:\\ 
   \Tab \Tab won-medals-of-in = \\
   \Tab \Tab \Tab \Tab
             PROJ[won-medals-of-in-1 = won-gold-in-1,
                       won-medals-of-in-3 = won-gold-in-2,\\
   \Tab \Tab \Tab \Tab \Tab \Tab \Tab \Tab 
                       won-medals-of-in-2 = Val(MedalKind,'G')] won-gold-in\\
   \Tab \Tab \Tab UNION\\
   \Tab \Tab \Tab \Tab
             PROJ[won-medals-of-in-1 = won-silver-in-1,
                       won-medals-of-in-3 = won-silver-in-2,\\
   \Tab \Tab \Tab \Tab \Tab \Tab \Tab \Tab   
                       won-medals-of-in-2 = Val(MedalKind,'S')] won-silver-in\\
   \Tab \Tab \Tab UNION\\
   \Tab \Tab \Tab \Tab 
             PROJ[won-medals-of-in-1 = won-bronze-in-1,
                       won-medals-of-in-3 = won-bronze-in-2,\\
   \Tab \Tab \Tab \Tab \Tab \Tab \Tab \Tab
                       won-medals-of-in-2 = Val(MedalKind,'B')] won-bronze-in\\
   \Tab \Tab MedalKind.code = \\
   \Tab \Tab \Tab \Tab \{   $\left< \SF{MedalKind.code-1 = Val(MedalKind,'G'), 
                                        MedalKind.code-2 = 'G'} \right>$,\\
   \Tab \Tab \Tab \Tab \Tab $\left< \SF{MedalKind.code-1 = Val(MedalKind,'G'), 
                                        MedalKind.code-2 = 'G'} \right>$,\\
   \Tab \Tab \Tab \Tab \Tab $\left< \SF{MedalKind.code-1 = Val(MedalKind,'G'), 
                                        MedalKind.code-2 = 'G'} \right>$ \}\\
   End Transformation schema.
\end{tabular}} \]
\SLabel{table}{ExTransformation}
\caption{The running example transformation}
\end{table}

An observant reader might now note that the transformation scheme
does not cater for the transformation of the constraints
defined over the types involved in the transformation.
In \cite{PhdThesis:89:Halpin:NIAMForm}, and 
\cite{Book:94:Halpin:ORM} transformation of constraints 
was covered by corollaries to the basic schema transformations.
While useful, this approach leads to many corollaries to deal with
different classes of constraints. Moreover, it 
provides no solution for constraints in general, and currently ignores
most textual (non graphical) constraints that
must be formulated in some formal textual language.
The approach taken in this article, is to continue enforcing the 
(uniqueness) constraints on the transformed relationships on the 
(now) derived relationships.
So in the Olympic games schema in schema (a), the \SF{won-gold-in}
relationship is derivable, while we enforce the 
uniqueness of its first role on this derived relationship.
Using a constraint re-writing mechanism, constraints on derived object types
can be re-written to constraints on the non derivable types.
Such a re-writing mechanism, however, very much depends on the language
used to express constraints and derivation rules.
In the next section we briefly return to this issue.

\subsection{Semantics of transformation schemes}
Although we do not provide a formal semantics of the language used to 
specify the transformation schemes, we do presume the
existence of three functions providing these semantics.
When an actual language is defined these functions will become concrete.
The (partial) functions are:
\[ \arraycolsep=0pt \begin{array}{rcl} 
   \From  &:&~ {\tt TransSchema}
         \Carth {\tt ParList} \PartFunc \Schema \\
   \To    &:&~ {\tt TransSchema}
         \Carth {\tt ParList} \PartFunc \Schema \\
   \Sch   &:&~ {\tt TransSchema}
         \Carth {\tt ParList} \PartFunc \Schema 
\end{array} \]
where ${\tt ParList} \subseteq ((\Preds \union \Types)^*)^*$
such that in each $X \in {\tt ParList}$ any role or type occurs
only once.
These functions are partial since some combinations of transformation
schemes and lists of parameters may define incorrect transformations.

The $\From$ function returns the schema components that will
be changed by the transformation.
What exactly is going to happen with these schema components depends on 
the aim with which the transformation is applied.
In the next section these aims will be discussed in more detail, together
with their different semantics.
The result of the $\From$ statement is given as a (sub) schema.
As this usually is a sub-schema without a proper context, this is
not likely to be a complete and correct ORM schema.
In our example the schema resulting from the $\From$ function
only contains the three relationship types from the (b) variation of the 
olympic games schemas, without the \SF{Country} and \SF{Quantity}
object types as such.

Similarly, the $\To$ function returns the added schema components
in the form of a sub-schema.
This returned schema is usually incomplete as well since it also
misses the proper context.
The $\To$ function not only returns the components
listed by the \SF{To} statement in the transformation scheme, but also 
the derivation and update rules.
In the schema resulting from the transformation, these rules are required 
to derive the instances of the transformed types and translate the updates 
of the transformed types to updates of the new types.

Finally, the $\Sch$ function yields all schema components listed in the
transformation scheme, and returns this as the schema.
However, the update rules are ignored.
The resulting schema provides the {\em context} of the schema transformation,
and is used as a base for providing equivalence preservation
proofs.

Whenever we use these functions we will apply the style of denotational
semantics (\cite{Book:77:Stoy:DenSem}). 
So we will for example write $\DenSem{\Sch}{t}{X} = \Schema_i$
to express the fact that $\Schema_i$ is the subschema that resuts
when applying function $\Sch$ to transformation scheme $t$ with parameter
list $X$.

\subsection{Inverse transformations}
If $t$ is an equivalence preserving schema transformation, then
$\SF{Inv}(t)$ denotes the inverse transformation.
In this case the lists provided by the \SF{To} and \SF{From}
statements have to be swapped, as well as the derivation
rules and update rules.
So if $t$ is a transformation, then $\DenSem{\Sch}{\SF{Inv}(t)}{X}$
has as derivation rules the update rules of $t$ (with the proper
parameters from $X$ instantiated), and vice versa.

In general it only makes sense to invert an equivalence preserving 
transformation scheme.
We will now take a closer look at the equivalence preservation 
properties which transformation schemes may possess.

\subsection{Properties of transformation schemes}
With respect to equivalence preservation, there are two 
transformation cases in which we are interested.
A transformation can be equivalence preserving or it can strengthen
an existing schema.
When a schema is strengthened, the number of correct populations decreases,
but this is sometimes seen as an acceptable trade-off to gain efficiency.

We generalise the above properties to transformation schemes.
A transformation scheme is equivalence preserving if and only if all correct
applications lead to equivalence preserving transformations.
Formally:
\[ \Al{\tuple{T,X} \in \FuncDom(\Sch)}{
          \DenSem{\Sch}{T}{X} \equiv \DenSem{\Sch}{\SF{Inv}(T)}{X}} \]
The example transformation scheme provided in \SRef{ExTransScheme} is 
equivalence preserving.

A transformation scheme is strengthening if and only if all correct applications
lead to strengthening transformations.
Formally:
\[ \Al{\tuple{T,X} \in \FuncDom(\Sch)}{
          \DenSem{\Sch}{T}{X} \Stronger \DenSem{\Sch}{\SF{Inv}(T)}{X}} \]
These last two properties can be proven by generalising the proof of a
concrete transformation.
How a concrete transformation can be proven to be equivalence preserving
or strengthening is discussed in \SRef{SchemaEquiv}.
For obvious reasons, the generalisation needs to be done to the parameters 
of the transformation scheme.

It is interesting to note that the derivation rules and update rules
provided with a transformation scheme are now used as the {\em conservative
extension} needed to prove the direct equivalence of the schema before
and after the transformation.

\subsection{Distributive updates}
Finally, we put one more restriction on the derivation and
update rules.
This restriction has as a benefit that the view update problem is
avoided.

When tranforming a conceptual schema $\Schema_i$ to a
data schema $\Schema_j$, the user will still want to perform the
updates as if they are done on the original conceptual schema.
That is why we have added the update rules.
These rules translate the updates of the conceptual types that
have been replaced by other types to updates of these replacement
types.
Conversely, when querying the database, the user is not interested in
the schema used for the actual storage of the data, but rather the
original conceptual schema.
In doing so, however, we may find ourselves exposed to the view update
problem.

To allow the user to specify updates on the conceptual level such that
they can be processed directly, we require
that the update rules are {\em update distributive}.
A set of update rules can be regarded as a function
    $\mu: \Populations \Func \Populations$ that takes a population
of the original schema and transforms that into a population of the
actually stored data schema.
The set $\Populations$ contains all possible populations, so
$\Populations = \ObjTypes \Func \Powerset(\UniDom)$, where $\UniDom$
is the set of possible instances of object types.
The derivation rules perform the opposite function $\mu^{-1}$, and for an 
equivalence preserving schema transformation this $\mu^{-1}$ is the
inverse function of $\mu$.

Before continuing we need the following generalisation of binary
operations on sets.
Let $p_1, p_2 \in \Populations$, then we can generalise each 
binary operation $\Theta$ on sets (of instances) to populations as
a whole by:
\[ (p_1 ~\Theta~ p_2)(x) = p_1(x) ~\Theta~ p_2(x) \]
A population transforming function: 
   $\mu: \Populations \Func \Populations$.
is called {\em update distributive} iff for
$\Theta \in \setje{\union,\SetMinus}$ and a correct
schema $\Schema$ we have:
\[ \IsPop(p,\Schema) \land \IsPop(p ~\Theta~ x,\Schema) \implies
   \mu(p ~\Theta~ x) = \mu(p) ~\Theta~ \mu(x) \]
This allows us to define a further restriction on the update rules specified
in a transformation scheme:
\begin{quote}
   Let $\mu$ be the population transformation function following from the
   update rules from a given transformation scheme $t$, then that $\mu$
   must be update distributive.
\end{quote}
If $t$ is equivalence preserving, then obviously the update rules of 
$\SF{Inv}(t)$ must define an update distributive function as well.
From the definition of $\SF{Inv}(t)$ follows that this function must
then necessarily be the inverse population transformation function 
following from the derivation rules in $t$.

With such a $\mu$ we can now safely translate any update of the
population of the original schema to an update of the transformed
schema.
For a conceptual schema optimisation process this means that a user
can continue specifying updates on the original conceptual schema.

Proving this property for a concrete transformation scheme is usually
not hard.
For the running example in this section it follows from the observation
that each tuple added to (or deleted from) the population of a 
relationship $f_i$ is turned into an addition (or deletion) of a tuple 
from the relationship $h$ with an extra column containing the unique 
constant $v_i$.
Conversely, when considering the inverse transformation, 
each update to the relation $h$ is translated to an update of one
of the relations $f_i$. 
Which one depends on the value of the $u$ column of the tuples involved
in the update.

Typical update rules that are now excluded due to this requirement are:
\begin{enumerate}
    \item rules containing aggregations like: summation, maximums, etc.
    \item rules containing encodings which need to consider the entire
          existing population to perform the encoding.
\end{enumerate}

   \section{Transformation Steps}
\SLabel{section}{ApplyTrans}

As stated before, in this article we are mainly interested in equivalence
preserving or strengthening transformations.
In the previous section we introduced a mechanism that enables us to
define transformation schemes which can be applied to a concrete
data schema.
Other schema transformations, like the ones used in the conceptual
schema design procedure, usually do not lend themselves to a representation
as a general transformation scheme.
The transformations in the conceptual schema design procedure have a
much more ad-hoc character.

When focusing on the equivalence (or strengthening) transformation
schemes, there are roughly three reasons to apply transformation schemes:
\begin{enumerate}
   \item to select an alternative conceptual schema which is regarded
         as a better representation of the universe of discourse,
   \item for the enrichment of the schema with derivable parts creating 
         diverse alternative views on the same conceptual schema as 
         a part of the original schema,
   \item to optimise a finished conceptual schema before mapping it to
         a logical design,
\end{enumerate}
The latter application is of course the main focus of this article.
Nevertheless, we will also discuss the other applications of the
schema transformations.

As an example, consider the transformation from \SRef{\HospitalA} to
\SRef{\HospitalB}.
One (the user community in conjunction with the modeller) might consider
the second schema to be a better conceptual representation of the universe 
of discourse.
On the other hand, one might decide to let both alternatives co-exist
together, this means that either \SRef{\HospitalA} or \SRef{\HospitalB}
can be used as a base for the implementation, and that users can
access the stored information in terms of the union of both schemas. 
Even more, once the conceptual schema is fixed, it may turn out
that the alternative chosen for the conceptual representation is
not the most optimal with respect to its implementation, in which case
the other alternative needs to be used as a base for the 
actual implementation.

Corresponding to these three ways to apply a transformation 
scheme, we have three differing semantics of a transformation scheme.
Before introducing these semantics, however, we first need to introduce a more 
elementary operation on schemas.

\subsection{Cleaning up a schema}
When transforming a given schema to a new schema, certain parts of 
the schema may become obsolete.
This means that these parts can (must) be removed from the new schema.
These removals may arise because:
\begin{enumerate}
   \item object types may have become isolated,
   \item derivation rules or update rules could 
         be collapsed into simpler rules, 
   \item constraints may have become derivable,
\end{enumerate}
We first focus on derivable types (together with their update
and derivation rules) that are removable from a
schema version $\Schema_i$.
A derivable type can be removed if:
\begin{enumerate}
   \item It has both an update rule and derivation rule that can be removed.

         Deleting a derivable type when the update or derivation rules
         cannot be removed would lead to a breach of the derivability of
         changed conceptual types.
 
   \item It is marked as an internal type.

         We do not want to remove conceptual types.

   \item When another type is dependent on this type.

         When no other type depends on the type for consideration of
         removal, then the removal of this type would lead to a collapse
         of a part of the information structure.
\end{enumerate}
Note that $2$ automatically follows from the fact that update rules can
only be defined for internal types.
These observations lead to the following formal
definition of the derivable types that can be removed ($R_i$):
\[ \begin{array}{llll}
   D_i & \Eq & \Set{\Defines(r)}{r \in \DerRules_i \land 
                             \Defines(r) \not\in \Depends(r)} 
         & \mbox{removable derivable types} \Eol
   U_i & \Eq & \Set{\Defines(r)}{r \in \UpdRules_i \land
                             \Defines(r) \not\in \Depends(r)} 
         & \mbox{removable internal types}\Eol
   P_i & \Eq & \Set{x}{\Ex{y \in \Types_i}{x \SubOf y}}
         & \mbox{types which have dependent types}\Eol
   R_i & \Eq & (D_i \intersect U_i) \SetMinus P_i 
         & \mbox{removable derivable types}
\end{array} \]
The second class of types that can be removed is the set of types that
are isolated, i.e. not connected to any other type.
This set is identified as:
\begin{eqnarray*}
   {\it UC}_i 
   & \Eq    & 
   \Set{x \in \Types_i}{\lnot \Ex{p \in \Preds_i}{\Player(p) = x}}
   \union  
   \Set{x \in \Types_i}{\lnot \Ex{y \in \Types_i}{y \SubOf x}}
\end{eqnarray*}
Now we know which types can be removed ($R_i \union {\it UC}_i$) 
in a schema version $\Schema_i$, we can define the sets of removable
derivation rules and update rules:
\begin{eqnarray*}
   RD_i & \Eq & \Set{r \in \DerRules_i}{\Defines(r) \in R_i \union {\it UC}_i}\\
   RU_i & \Eq & \Set{r \in \UpdRules_i}{\Defines(r) \in R_i \union {\it UC}_i}
\end{eqnarray*}
The actual $\CleanUp$ operation is defined recursively.
For a given schema version $\Schema_i$, the one-step cleanup
operation $\CleanUp^1(\Schema_i)$ leads to a new schema $\Schema_j$
where:
\begin{enumerate}
   \item $\Types_j = \Types_i \SetMinus {\it UC}_i \SetMinus R_i$

   \item $\ITypes_j = \ITypes_i \intersect \Types_j$

   \item $\DerRules_{j} = (\DerRules_i \SetMinus {\it RD}_i)|^{{\it RD}_i}$
         where $X|^{{\it RD}_i}$ replaces all references to rules in 
         {\it RD} by the body of these rules (i.e. substituting the 
         definition of the removed derivation rules in the remaining 
         rules).

   \item $\UpdRules_j = 
          (\UpdRules_i \SetMinus {\it RU}_i)|^{{\it RU}_i \union {\it RD}_i}$

         Note that in this case also the derivation rules that are removed
         need to be substituted as the update rules may refer to
         derivable types that have just been removed.

   \item $\Constraints_{j} = 
          \SF{Reduce}(\Schema_i,\Constraints_i|^{{\it RD}_i})$ 
         where \SF{Reduce} tries to remove derivable constraints and 
         rewrite the constraints to constraints on non-derivable types.
         Below we elaborate more on this function.
        
   \item $\Dom_{j} = \Set{\tuple{v,d} \in \Dom_i}{v \in \ValueTypes_j}$
\end{enumerate}
This defines the one-step clean-up operation.
After one clean-up step additional types may have become isolated.
Therefore, the clean-up operation needs to be applied
recursively.
For $n > 1$ we therefore have:
\[ \CleanUp^n(\Schema_i) = \CleanUp^1(\CleanUp^{n-1}(\Schema_i)) \]
From this the general cleaning up function can be defined by:
\[ \CleanUp(\Schema_i) = \CleanUp^n(\Schema_i) \mbox{~~where $n$ is such that:~} 
   \CleanUp^n(\Schema_i) = \CleanUp^{n+1}(\Schema_i) \]
This leaves the definition of the \SF{Reduce} function. 
The \SF{Reduce} function is needed for efficiency reasons,
since it is much more efficient to enforce constraints on
base types (and eventually on tables) then on derived
types.
Furthermore, algorithms that map data schemas to internal
schemas (e.g.\ a relational schema) typically make mapping
decisions based on the constraints that hold on the base
types.
However, the exact definition of such a function depends very much
on the language chosen for the specificiation of constraints
and derivation rules.

For example in the transformation given in \SRef{\Olympics}, 
one would like to have the system derive the uniqueness 
constraints on the three binary fact types in the resulting schema 
from the single two-role uniqueness constraint in the original schema.
Given a formal language for constraint specification and derivation
rule specification (for example a relation algebra), one could specify
a set of re-write rules for such constraint and derivation rule
specifications.

\subsection{Schema alternatives}
The first semantic interpretation of a transformation scheme we
consider leads to schema alternatives.
These semantics are provided by the function:
\[ \Alternative: {\tt TransSchema} \Carth {\tt ParList} \Carth \Schema \PartFunc \Schema \]
which is defined by
\[ \DenSem{\Alternative}{T}{X,\Schema_i} ~\Eq~ \CleanUp(\Schema_j) \]
where (using $\Schema_f = \DenSem{\From}{T}{X})$ and
             $\Schema_t = \DenSem{\To}{T}{X}$)
the components of $\Schema_j$ are defined as:
\begin{enumerate}
   \item $\Types_j  = \Types_i  \union \Types_t$.

   \item $\ITypes_j = \ITypes_i \union \Types_f$. \label{IntTypes}

         We cannot simply remove the changed types ($\Types_f$) 
         since some of them may be used in constraints for the 
         derivation or construction of other types.
         Therefore they (initially) need to remain present in the 
         schema, unless a closer study reveals that a type is
         indeed not needed ($\CleanUp$).
          
   \item $\DerRules_j = \DerRules_i \union \DerRules_t$.

   \item $\UpdRules_j = \UpdRules_i \union \UpdRules_t$.

   \item $\Constraints_j = \Constraints_i \union \Constraints_t$

   \item $\Dom_j = \Dom_i \union \Dom_t$.
\end{enumerate}
The schema transformation must, however, obey certain rules.
If these rules are violated, the schema transformation is considered
undefined.
The rules are:
\begin{enumerate}
   \item The correctness of schemata is preserved by the transformation: 
         \[ \IsSch(\Schema_i) \implies \IsSch(\CleanUp(\Schema_j)) \]
   \item The types changed by the transformation are present in the
         original schema as non-derivable types:
         $\Types_f \subseteq \Types_i \SetMinus 
          \Set{\Defines(r)}{r \in \DerRules_i}$.
   \item The new types should not already be present in the current
         data schema:
         $\Types_t \intersect \Types_i = \emptyset$.
\end{enumerate}

\subsection{Schema enrichment}
The second semantic interpretation is the enrichment of an
existing data schema.
No types are removed, but the set of known types is enriched
by new types defined in terms of the existing ones.
The semantics are provided by the function 
\[ \Enrich: {\tt TransSchema} \Carth {\tt ParList} \Carth \Schema \PartFunc \Schema \]
which is defined by
\[ \DenSem{\Enrich}{T}{X,\Schema_i} ~\Eq~ \CleanUp(\Schema_j) \]
where $\Schema_j$ is defined the same as for $\Alternative$
except for the internal types:
\begin{enumerate}
   \item[\ref{IntTypes}]
        The internal types of the schema remain unchanged:
        $\ITypes_j = \ITypes_i$.
\end{enumerate}
The same extra conditions as defined on $\Alternative$ apply
for $\Enrich$ as well.

\subsection{Schema optimisation}
The third interpretation defines a transformation leading to a
data schema optimised for internal representation:
The semantics are provided by the function 
\[ \Optimise: {\tt TransSchema} \Carth {\tt ParList} \Carth \Schema \PartFunc \Schema \]
which is defined by
\[ \DenSem{\Optimise}{T}{X,\Schema_i} ~\Eq~ \CleanUp(\Schema_j) \]
where $\Schema_j$ is defined the same as for $\Alternative$
except, again, for the internal types:
\begin{enumerate}
   \item[\ref{IntTypes}]
        The internal types of the new schema are:
        $\ITypes_j = \ITypes_i \union~ (\Types_t \SetMinus \Types_i)$.

        All types added are for internal purposes only.
\end{enumerate}
The same extra conditions as defined on $\Alternative$ apply
for $\Optimise$ as well.

   \section{Conclusions and Further Research}
\SLabel{section}{Concl}

In this article we have presented an approach to automated conceptual schema
transformations.
We have sketched a broader context for these transformations, and have 
focussed the attention on transformations improving the conceptual quality 
of schemas as well as transformations leading to more efficient 
implementations.

Given a concrete language for constraint and derivation rule specifications,
a re-write system must now be developed for this language that allows
constraints to be re-written (as much as possible) in terms of
base types.
It is planned to implement the ideas presented in this article in
the context of a project aiming for the development of a generic conceptual
(data) modelling CASE Tool (\cite{Report:95:Proper:CDMKernel}).
Furthermore, the pool of strengthening and equivalence 
preserving schema transformation schemes needs to be extended.

Heuristics and algorithms to drive a schema optimisation process have been 
developed \cite{Book:94:Halpin:ORM}, but these need to be extended to cover more 
cases (e.g.\ use of complex types).
One cannot expect a user to manually select both the schema components and
a transformation scheme to be used in a schema transformation step.
A tool for performing schema optimisation in a 
semi-automated way is currently being designed.

   \BIBLIOGRAPHY{alpha}
\end{document}